\documentclass[12pt]{article}
\usepackage{amsmath}
\usepackage{amsfonts}
\usepackage{amssymb}
\usepackage{mathrsfs}
\usepackage{color}
\usepackage[utf8]{inputenc}
\usepackage{doi}
\usepackage{url}
\pdfoutput=1 

\usepackage[square,numbers]{natbib}
\usepackage[T1]{fontenc}

\usepackage{hyperref}

\usepackage{frcursive}
\usepackage{pdfpages}
\usepackage{appendix}
\usepackage{graphicx}
\usepackage{natbib}

\usepackage{dsfont}
\usepackage{bbold}
\usepackage{bbm}

\newtheorem{proposition}{Proposition}[section]

\newenvironment{proof}[1][Proof]{\noindent\textbf{#1.}}{\ \rule{0.5em}{0.5em}}

\def\ds{\displaystyle}

\begin{document}

%------------------------------------------------------------------------
\begin{center}
{\Large
{\sc Modeling space-time trends and dependence in extreme precipitations of Burkina Faso  by the approach of the  Peaks-Over-Threshold}
}

\bigskip
B\'ewentaor\'e Sawadogo$^{1,2}$  \& Diakarya Barro$^{1,3}$
\bigskip

{\it
$^{1}$ LANIBIO, Université Joseph KI-ZERBO, BP: 7021, Ouagadougou 03, Burkina Faso, sbewentaore@yahoo.fr\\
$^{2}$   Université Paris Saclay, INRAe, AgroParisTech, UMR MIA-Paris, 75005, Paris, France, sbewentaore@yahoo.fr\\

$^{3}$ UFR-SEG, Université Thomas SANKARA, BP: 417 Ouagadougou 12, Burkina Faso, dbarro2@gmail.com
}
\end{center}
\bigskip

%------------------------------------------------------------------------
\textbf{Abstract.}
Modeling  extremes of climate variables in the framework of climate change is a particularly difficult task, since it implies taking into account spatio-temporal non-stationarities.
In this paper, we propose a new method for estimating extreme precipitation at the points where we have not observations using information from marginal distributions and dependence structure. To reach this goal we combine two statistical approaches of extreme values theory allowing on the one hand to control temporal and spatial non-stationarities via a tail trend function with a spatio-temporal structure in the marginal distributions and by modeling on the other hand the dependence structure by a latent spatial process using generalized $\ell$-Pareto processes. This new methodology for trend analysis of extreme events is applied to rainfall data from Burkina Faso. We show that extreme precipitation is spatially and temporally correlated for distances of approximately 200 km. Locally, extreme rainfall has more of an upward than downward trend.\vspace{0.1cm}

\medskip

\section{Introduction}
In the framework of climate change, the modeling and accurate prediction of the magnitude and extent of extreme events that occur in space and time of climate variables is a particularly difficult task, since it implies taking into account spatial and temporal non-stationarities. Nowadays there is a general consensus in the scientific community that climate change has accelerated in recent decades and that the climate will continue to change in the coming decades, mainly due to natural and anthropogenic changes (IPCC 2007, 2018,2019). This change is manifested in most regions of the world by a resurgence of heavy rainfall, heat waves and pollution peaks with very significant economic and social consequences. From a statistical point of view, the hypothesis of stationarity becomes untenable and debatable, and thus raises the question of detecting trends in extreme events, which has been the subject of much attention over the last 30 years. The classical Theory of Extreme Values extended both to non-stationary and non-independent observations provides a rigorous mathematical framework to deal with this question of trend detection in extremes (\cite{einmahl2016statistics}, \cite{ ferreira2017estimating},  \cite{mefleh2020trend}, \cite{cabral2020space}).\vspace{0.3cm}

This issue was first generally studied in extreme value literature by parametric pointwise approaches in which an extreme value model (GEV or GPD) is fitted to the data at each site in turn, leaving the parameters of the marginal distributions to evolve with time or other significant covariates(\cite{davison1990models}, \cite{olsen1998risk},  \cite{parey2010different}, \cite{cooley2013return}, \cite{salas2014revisiting}).  Although it is relatively simple to construct non-stationary models in the univariate framework, it is more difficult to account for spatial and temporal trends in these univariate models. The spatial and temporal dimension of extreme events was first developed for block maxima in a stationary framework and modeled by max-stable processes (\cite{de1984spectral}, \cite{hann2006extreme}, \cite{bechler2015conditional}, \cite{diakarya2012spatial, diakarya2017asymptotic}). These spatial models have been readapted to handle non-stationarities  induced by global warming (\cite{huser2016non},  \cite{chevalier2020modeling}). Although attractive, these models are expensive in computing time when tending towards large dimensions and are not adapted for modeling threshold exceedances.\vspace{0.3cm}

The threshold exceedance approach introduced by \cite{balkema1974residual} and \cite{pickands1975statistical} was first extended to multivariate environments (\cite{rootzen2018bmultivariate}) before being generalized to functional data (\cite{ferreira2014generalized}, \cite{dombry2015functional}, \cite{DeFondeville2018, DeFondeville2020}) to give birth to the family of generalized $\ell$-Pareto processes.  This family of processes  are better adapted to model the spatial dependence structure of exceedances data. However, these approaches do not sufficiently take into account marginal non-stationarities and spatial dependence between margins. In this paper we propose a new method to capture non-stationarities in marginal distributions, while taking into account the spatial dependence structure. To reach this goal we combine two statistical approaches of extreme value theory allowing on the one hand to control temporal and spatial non-stationarities via a tail trend function with a spatio-temporal structure in marginal distributions(\cite{einmahl2016statistics}, \cite{ferreira2017estimating}, \cite{cabral2020space}) and by modeling on the other hand the dependence structure by a hidden stationary auxiliary process using generalized $\ell$-Pareto processes \cite{DeFondeville2020}. \vspace{0.3cm}

The article is organized as follows: section 2 provides a brief overview on generalized $\ell$-Pareto processes. Section 3 details the methodology implemented and the methods used to estimate the parameters. In section 4 we present our main results, and in particular we compare the stationary and  non-stationary return levels computed from the developed approaches.

\section{Overview on generalized $\ell$-Pareto process}\label{secpareto}
Let $S$ be a compact subset of $\mathbb{R}^{d}$ and $T$ be a compact subset of $\mathbb{R}^{+}$ denoting the spatial and temporal domain respectively. We note by $\mathcal{C}(S)$ the set of continuous real functions on $S$ with the uniform norm $\parallel. \parallel_{\infty}$; $\mathcal{C}_{+}(S)$ its restriction to non-negative functions deprived of the null function and  $\mathscr{M}\left\lbrace \mathcal{C}_{+}(S) \right\rbrace $ the class of measures associated with $\mathcal{C}_{+}(S)$. \vspace{0.1cm}

Let $\left\lbrace Z(s), s\in S\right\rbrace$ be a  latent spatial stochastic process indexed by $s\in S\subset \mathbb{R}^{d}$ with sample paths in $\mathcal{C}(S)$ of continuous  marginal distribution $F_{Z}$ and common right endpoint $z_{F}$. As in (\cite{ferreira2014generalized},  \cite{DeFondeville2020}), we assume that the stationary process $Z$ is a  general functional regular variation, noted $Z\in GRV(\gamma, a_n, b_n, \Lambda)$, i.e that there exists suitable sequences of continuous  functions $a_{n}:S\longrightarrow\mathbb{R}^{+}$, $b_{n}:S\longrightarrow \mathbb{R}$ and $\gamma\in\mathbb{R}$ such that 
\begin{eqnarray}\label{eqhypoGRV}
nP\left[\left(1+\gamma\left((Z-b_{n})/a_{n}\right)  \right)^{1/\gamma}_{+} \in . \right] \longrightarrow \Lambda(.), n\longrightarrow +\infty,
\end{eqnarray} 
where $\Lambda$ is a non-zero measure in $\mathscr{M}\left\lbrace \mathcal{C}_{+}(S)\right\rbrace$ and homogeneous of order $-1$, with $\Lambda\left(tA \right)=t^{-1}\Lambda\left(A\right)$ for any positive real $t>0$ and Borel set $A\subset \mathcal{C}_{+}(S)$. \vspace{0.1cm}

In the multivariate and spatial framework a threshold exceedances for a random function $Z=\left\lbrace Z(s), s\in S\right\rbrace$ is  defined by Dombry and Ribatet \cite{dombry2015functional} to be an event of the form $\left\lbrace \ell(Z)> u \right\rbrace$ for some $u\geq 0$, where $\ell: \mathcal{C}(S)\longrightarrow \mathbb{R}^{+}$ is a continuous and homogeneous non-negative risk function, i.e, there exists $\alpha >0$ such that $\ell(\lambda y)=\lambda^{\alpha}\ell(y)$ when $y\in \mathcal{C}_{+}(S)$ and $\lambda>0$. The risk function $\ell$ determines the type of extreme events of interest. For example, such a function can be the maximum, minimum, average or value at a specific point $s_{0}\in S$.  Moreover, as in (\cite{DeFondeville2020}, \cite{Engelke2019}) we assume that there exists a continuous and positive real function $A_{Z}$  such that the sequence $a_{n}$ and the risk function $\ell$ satisfy the following asymptotic decomposition
\begin{eqnarray} \label{eqhypoasymptotic}
\begin{array}{c|c|c}
\displaystyle\lim_{n\rightarrow\infty}\sup_{s\in\mathcal{S}}&\dfrac{a_{n}(s)}{\ell(a_{n})}-A_{Z}(s)&= 0, ~~~~ \text{i.e} ~~a_{n}(s)\approx \ell(a_{n}) A_{Z}(s),~~ n\rightarrow\infty
\end{array} 
\end{eqnarray}
The marginal distributions of $Z$ are supposed to belong to a location-scale family, thus ensuring  a constant $\gamma\in\mathbb{R}$ shape parameter over any $S$, i.e. that there exist a $H$ distribution function such that
\begin{eqnarray}
P\left(Z(s)\leq z \right)= H\left[\left( z(s)-B_{Z}(s)\right)/A_{Z}(s) \right],
\end{eqnarray} 
where $A_Z: S \rightarrow \left(0; \infty\right[$ verifies asymptotic decomposition(\ref{eqhypoasymptotic})  and $B_Z : S \rightarrow \mathbb{R}$ are continuous functions. In particular $H$ must belong to the domain of attraction of a generalized extreme value (GEV) distribution of index $\gamma\in\mathbb{R}$, i.e.,  $F_{Z}\in \mathscr{D}(G_{\gamma})$. In other words, there are appropriate normalization real sequences  $\tilde{a}_{n}>0$, $\tilde{b}_{n} \in \mathbb{R}$ such that  $\ds\lim_{n\rightarrow \infty}H^{n}\left(\tilde{a}_{n}z+\tilde{b}_{n}\right)=G_{\gamma}(z)$. The condition of the max-domain of attraction is equivalent to 
\begin{eqnarray}\label{eqDomGEV}
\ds\lim_{n\rightarrow +\infty} n\left[ 1-H\left( \tilde{a}_{n}z+ \tilde{b}_{n} \right)\right]=-\log G_{\gamma}(z), ~~ z>0, 
\end{eqnarray}
Under these conditions and the assumption of the general functional regular variation the suitable sequence of continuous functions $a_{n}(s)$ and $b_{n}(s)$, satisfy: 
\begin{eqnarray}\label{eqChoix_an_and_bn}
a_{n}(s)=\tilde{a}_{n}A_{Z}(s), ~~~~~~ b_{n}(s)=B_{Z}(s)+A_{Z}(s)\tilde{b}_{n},~~~~ s\in S.
\end{eqnarray}
Functions $A_{Z}(s)$ and $B_{Z}(s)$ as well as the spatial dependence structure of $Z$ are assumed to belong to a family of parametric functions $\left\lbrace A_{Z,\theta_{A}}: \theta_{A}\in\Theta_{A} \right\rbrace $ ; $\left\lbrace B_{Z,\theta_{B}}: \theta_{B}\in\Theta_{B} \right\rbrace $ and $ \left\lbrace \Lambda_{\theta_{\Lambda}}: \theta_{\Lambda}\in\Theta_{\Lambda} \right\rbrace$.  Under minimal assumptions on the risk function and assumptions  (\ref{eqhypoGRV},  \ref{eqhypoasymptotic}) of $Z$, the conditional distribution of $\ell$-exceedance for some threshold  $u\geq0$  of the  process  $(Z-b_n)/\ell(a_n)$ can be approximated by a generalized $\ell$-Pareto process,  for $n$ large enough(\cite{DeFondeville2020})
\begin{eqnarray}\label{eqLPareto}
\displaystyle P\left\lbrace  \Bigg \lfloor\dfrac{ Z-b_{n}}{\ell\left( a_{n}\right) } \Bigg \rfloor \in A \mid \ell\left(\dfrac{ Z-b_{n}}{\ell\left( a_{n}\right) }\right) \geq u \right\rbrace \longrightarrow P\left\lbrace W_{\ell}\in A\right\rbrace, n\longrightarrow \infty,  
\end{eqnarray}
where $ \lfloor z \rfloor =\max\left(z,A_{Z}\gamma^{-1}\right)$ if $\gamma >0$ and $ \lfloor z  \rfloor =z$ if $\gamma\leq 0$.  $W_{\ell}$ is a non-degenerate stochastic process over S and belongs to the family of generalized $\ell$-Pareto processes with tail index $\gamma$, zero location, scaling function $A_{Z}$ and  limit measure $\Lambda$.
Specifically,  a generalized  $\ell$-Pareto process  $W_{\ell}$  associated to the limit measure $\Lambda$ and tail index $\gamma \in\mathbb{R}$ is a  stochastic process taking values in $\left\lbrace z\in \mathcal{C}_{+}(S): \ell\left\lbrace (z-b)/\ell(a)\right\rbrace\geq 0 \right\rbrace$  and  defined by: 
\begin{eqnarray}\label{eqDefLpareto}
W_{\ell}=\left\{\begin{array}{rl}
a\left(Y_{\ell}^{\gamma}-1\right) /\gamma  +b, ~ \gamma \neq 0\\
a\log Y_{\ell} +b,~~~~~~~~\gamma=0 ,
\end{array} 
\right.;
\end{eqnarray}
where $a= \ell(a)A_{Z}>0$ and $b$  are  continuous functions on $S$  respectively scale and location functions and $Y_{\ell}$ is a stochastic process whose probability measure is completely determined by the limit measure $\Lambda$.  For the modelling of spatial dependence structure of latent process, we use $Y_{\ell}$ whose margins are in the Frechet max-domain of attraction with tail index $\gamma=1$, as the process of reference. 
A  pseudo-polar decomposition of $Y_{\ell}$ in (\ref{eqDefLpareto}) leads to the following formulation (\cite{dombry2015functional}, \cite{DeFondeville2018})
\begin{eqnarray}\label{eqDecopoPseudoPolar}
Y_{\ell}=RQ
\end{eqnarray}
where $R$ is a unit Pareto random variable of index $\gamma_{R}$ representing the intensity of process, and $Q$ is stochastic process denoted the angular component with state space  $S$ and taking values in  $\mathscr{S} =\left\lbrace y\in\mathcal{C}_{+}\left(S \right): \parallel y\parallel_{1}=1  \right\rbrace $ whose probability measure is characterized by limit measure $\Lambda$. More details on generalized $\ell$-Pareto processes can be found in (\cite{ferreira2014generalized}, \cite{dombry2015functional}, \cite{DeFondeville2018, DeFondeville2020}). 

\section{Space-time trends detection}
Let $X=\left\lbrace X_{t}(s),s\in S,t\in T \right\rbrace$ be a continuous non-stationary space-time stochastic process with sample paths in the family of continuous functions $\mathcal{C}(S\times T)$ equipped with the uniform norm $\parallel. \parallel_{\infty}$, where 
$S\times T\subset \mathbb{R}^{d}\times\mathbb{R}^{+}$ and $\mathcal{C}_{+}(S\times T)$ its restriction to non-negative functions deprived of the null function.  In practice $X$ is observed at each stations $s_{1},\cdots,s_{m}$ and at given dates $t=1, \cdots, n$. Let $ F_ {t, s} $ be the  continuous univariate marginal distribution with a common right endpoint $ x_ {F} $ and $ Z = \left\lbrace Z (s), s \in S \right \rbrace $ an unobserved latent spatial stochastic process with sample paths in $\mathcal{C}(S\times T)$ satisfying the properties described in the section (\ref{secpareto}) and the proportional tail condition such that 
\begin{eqnarray} \label{eqLienXandZ}
\displaystyle\lim_{x\rightarrow x_{F}}\dfrac{P\left(X_{t}(s) > x \right) }{P\left( Z(s) >x \right) } =c_{\theta}\left(\frac{t}{n},s\right), 
\end{eqnarray}
where  $c_{\theta}: \left[ 0,1\right] \times S\longrightarrow \left( 0, \infty \right[$ is an assumed continuous and positive function depending on a parameter vector $\theta\in \Theta\subset\mathbb{R}$, called tail scale function or skedasis function (\cite{einmahl2016statistics},  \cite{ferreira2014generalized, ferreira2017estimating}, \cite{cabral2020space}). In added we assume that the continuous marginal distributions $F_{Z}$ have a some common right endpoint as $F_{t,s}$. The skedasis function describes the evolution of extreme events jointly in space and time.  The tail trend function $c_{\theta}$ is designed to ensure  uniqueness of $c_{\theta}$ such that:
\begin{eqnarray}\label{eqUnicite_de_c}
\dfrac{1}{m}\sum_{j=1}^{m}\int_{0}^{1}c_{\theta}\left( u,s_{j}\right)du =1, ~~ u\in\left[0,1 \right],~~ s_{j}\in S .
\end{eqnarray}
In the framework of the model (\ref{eqLienXandZ}), Mefleh et al.\cite{mefleh2020trend} shows that the empirical point measure converges  in distribution in the space of point measure $\mathcal{M}_{p}=\mathcal{M}_{p}\left([0,1]\times (0,\infty]\right)$ to a Poisson point process with intensity measure $c(u)du\gamma z^{-(\gamma+1)}dz$ on $[0,1]\times (0,\infty]$. In such a case the times of  exceedances for high threshold $x$ and the value of exceedances are asymptotically independent with distributions respectively equal to the trend density function $c_{\theta}(u,s), s\in S$ and the Pareto distribution of tail index $\gamma$. 
Several models are eligible to model the function $c_{\theta}$, but in this study we opt for parametric models because of their flexibility in trend detection and in order to make extrapolations of the trend beyond the observed data. We are interested in monotonic log-linear and simple linear trend models of $c_{\theta}$.  We introduce a spatial structure in the  tail function by letting the parameter  $\theta$ evolve as a function of geographic coordinates (longitude, latitude), that is, 
\begin{eqnarray}\label{eqTrendsFunction}
\left\{\begin{array}{rl}
c_{\theta}^{1}(\frac{t}{n},s)=\dfrac{\theta(s)}{\exp\{\theta(s)\}-1}\exp\{\theta(s)\frac{t}{n}\}, ~~ \theta(s) \in \mathbb{R}; ~\frac{t}{n}\in \left[0, 1\right]\\\\
c_{\theta}^{2}(\frac{t}{n},s)=\theta(s)\left\lbrace 2\frac{t}{n}-1\right\rbrace +1, ~~ \theta(s)\in \left]-1; 1\right[; ~~\frac{t}{n}\in \left[0, 1\right]
\end{array} 
\right.; 
\end{eqnarray}
where $\theta(s)=\theta_{0} +\sum_{k=1}^{2}\theta_{k}Y_{k}(s)$ with $(Y_{1},Y_{2})=(Long,Lat)$. The parameter vector $\theta=\left( \theta_{0},\theta_{1},\theta_{2}\right)$ is then estimated using the maximum likelihood method and multiple linear regression. 

\subsection{Marginal model of latent process}\label{secMarginal}
The marginal parameters $\gamma$, $a_{n}$, $A_{Z}$, $b_{n}$  and $B_{Z}$ of the  latent  spatial  process $Z$ are estimated  under the constraint that $u_n=\ell(b_n)+u\ell(a_n)$, following the modeling assumptions described in section(\ref{secpareto}). For simplicity purpose we choose $u=0$ so that $\ell$-exceedances of $Z$ are defined as events for which $\ell(Z)\geq\ell(b_n)$. In this case a natural choice of $u_n=\ell(b_n)$ is a high quantile of the variable $\ell(Z)$, for example $u_n=q_{0.95}\left\lbrace \ell\left(Z \right)\right\rbrace$.  In general, a parametric model may be necessary for $a_{n}$ and $b_{n}$, as in Engelke et al.\cite{Engelke2019}, but we consider in this work that $a_{n}(s_{j})=a_{j}>0$ and $b_{n}(s_{j})=b_{j}\in\mathbb{R}$  for any $j=1,\cdots, m$. The threshold stability of the generalized Pareto distributions does not allow us to identify the $b_{n}$ function without additional assumptions, so under the assumption that $\ell(B_Z)=0$, we set
\begin{eqnarray}
b_{n}(s)=u_{q'}\left\lbrace Z(s)\right\rbrace-\tilde{b}_{n}, ~~~ s\in S,
\end{eqnarray} 
where $\tilde{b}_{n}=\ell(b_{n})=u_n$ (see $\ref{eqhypoasymptotic}$ and $\ref{eqChoix_an_and_bn}$) and  $u_{q'}\left\lbrace Z(s)\right\rbrace$ is an empirical quantile of the order $q'$ of the $\ell$-exceedances at each location $s$, where $q'$ is chosen such that $\ell(b_{n})=\tilde{b}_{n}$ in order to impose the identifiability of the parameters. Thus, the tail index $\gamma \in\mathbb{R}$ and the scale parameters $a_{n}(s) > 0$ are estimated by maximizing the independent log-likelihood; that is  
\begin{eqnarray}\label{eqLogVraisIndep}
\ell_{indep}(\gamma,a_{n}(s_{j}))=\displaystyle\sum_{t=1}^{n}\sum_{j=1}^{m}\mathbb{1}_{\left\lbrace x_{t}(s_{j}) \geq b_{n}(s_{j})\right\rbrace }\log\left[\dfrac{1}{a_{n}(s_{j})}\left\lbrace 1+\gamma\dfrac{x_{t}(s_{j})-b_{n}(s_{j})}{a_{n}(s_{j})} \right\rbrace^{-1/\gamma -1}_{+} \right]
\end{eqnarray}
The estimate of parameters $A_{Z}$ and $B_{Z}$ are deducted from (\ref{eqChoix_an_and_bn}) under the assumptions of parameter identifiability, i.e, $\ell(A_{Z})=1$ and $\ell(B_{Z})=0$, which implies that $\ell(a_{n})=\tilde{a}_{n}$. Thanks to the equation(\ref{eqLienXandZ}) and the convergence of $Z$ exceedances to a GPD distribution we deduce a sample of $\left\lbrace Z_{t}(s)\right\rbrace$ from observations of $\left\lbrace X_{t}(s)\right\rbrace$ (\cite{ferreira2017estimating} and \cite{cabral2020space}) in the following manner:
\begin{eqnarray}\label{eqzobs}
\widehat{Z}_{t}(s)=\left\lbrace \widehat{c}_{\theta}\left(\frac{t}{n},s \right)  \right\rbrace^{-\widehat{\gamma}} \left[ X_{t}\left(s \right)-\frac{\left\lbrace \widehat{c}_{\theta}\left(\frac{t}{n},s\right)  \right\rbrace^{\widehat{\gamma}}-1 }{\widehat{\gamma}}\left(\widehat{\tilde{a}}_{n}-\widehat{\gamma}\widehat{\tilde{b}}_{n} \right)  \right],
\end{eqnarray}
where $\widehat{\gamma}, \widehat{\tilde{a}}_{n}, \widehat{\tilde{b}}_{n}$ and  $\widehat{c}_{\theta}$ are respectively consistent estimators of $\gamma, \tilde{a}_{n}, \tilde{b}_{n}$ and $c_{\theta}$ described in section (\ref{secMarginal}).

\subsection{Spatial extremal  dependence}
 After removing the non-stationarity the modeling is focused on the evaluation of the remaining extreme spatial dependence structure in $Z$. Thus we approach the limit distribution of $\ell$-exceedances of $Z$ by a generalized $\ell$-Pareto process \cite{DeFondeville2020}.  In the framework of the Brown-Resnick model, the angular component of $Q$ is a log-Gaussian random function whose underlying Gaussian process has stationary increments, which allows in particular to use classical dependence structures from the geostatistical literature to characterize the spatial dependence.  In order to better capture this dependence structure, we use a flexible parametric semi-variogram belonging to the class of power semi-variograms.  We estimate the parameter vector $\theta_{\Lambda}$ of the dependence structure using the gradient score method  or the censored likelihood method \cite{DeFondeville2018}.

\subsection{Non-stationary return period and return level }
The concept of  return period and level  becomes very ambiguous when we leave the stationary context to the non-stationary framework. In this paper, we have chosen to follow  return period based approaches,i.e., Expected Number of Exceedances(ENE) \cite{parey2010different} and Expected Waiting Time (EWT) (\cite{olsen1998risk}, \cite{cooley2013return}). In the ENE approach, the number of times the  variable $X_{t}(s), s\in S$ exceeds the return level value $x_{m}$ in  $m$  years is defined by $N_{m}=\sum_{t=1}^{n_{x}m}\mathbb{1}_{\left\lbrace X_{t}(s)>x_{m}, s\in S \right\rbrace}$ under non stationary context. The return level $x_{m}$ can be defined as the  value for which the expected number  of events exceeding $x_{m}$ in $m$ years equals to one, i.e., the return level $x_{m}$ is the solution of  the following equation:
\begin{eqnarray}\label{eqENE}
\ds 1=\sum_{t=1}^{n_{x}m}\left\lbrace 1- F_{t,s}\left(x_{m}\mid \theta_{t}(s) \right) \right\rbrace, ~~s\in S, 
\end{eqnarray}
where $n_{x}$ is the number of days in the year and $\theta_{t}(s)$ the vector of time-dependent marginal parameters or other covariates. Parey et al.\cite{parey2010different} uses the ENE method in a pointwise POT model where the parameters of the distribution of exceedances and the intensity of extreme event occurrences are described as polynomial functions of time. \vspace{0.3cm}

The EWT method was first proposed by  Olsen et al.\cite{olsen1998risk}, and then  derived by Salas and Obeysekera \cite{salas2014revisiting} using a geometric distribution with time-varying parameters. Under non stationary conditions, the  distribution describing waiting time $Y$ before the
first occurrence of an event exceeding the return level $x_{m}$ is
\begin{eqnarray}\label{eqgeometricNonStatCDF}
f_{s}(y)=P\left(Y(s)=y \right)=p_{y,s}\prod_{i=1}^{y-1}\left(1-p_{i,s} \right), y=1,2,\cdots, y_{\max}~~\text{and}~ s\in S,
\end{eqnarray}
where variable $Y$ is the day of the first occurrence of an event exceeding the  quantile $x_m$ , $p_{t,s}=1-F_{t,s}(x_{m}\mid\theta_{t}(s))$ is daily exceedance probability varying with time step t. $y_{\max}$ is the time when the daily exceedance probability $p_{t,s}$ is equal to 1 for an increasing-trend series or is equal to 0 for a decreasing-trend series.  Reused and simplified by \cite{cooley2013return}, the EWT approach defines the $m$-year return level $x_{m}$, as the  value for which the expected waiting time until an exceedance of this level is $m$ years, i.e,  $x_{m}$ is the solution of the equation: 
\begin{eqnarray}\label{eqEWT}
n_{x}m=E\left[ Y\right] =1+\ds\sum_{y=1}^{\infty}\prod_{t=1}^{y}F_{t,s}\left(x_{m} \mid \theta_{t}(s) \right), s\in S.
\end{eqnarray}
Using the relationship (\ref{eqLienXandZ}), $P(X_{t}(s)> x_{m})$ can be rewritten  $c_{\theta}(\dfrac{t}{n}, s)P(Z(s)>u)P(Z(s)>x_{m}\mid Z(s)>u)$ for $u<x_{m}$ and we obtain  the following results
\begin{proposition}\label{NewPropReturnLevel}
Let $ \{X_{t}(s),s\in S, t\in T\} $ be a non-stationary stochastic process defined on a region $S\subset\mathbb{R}^{d}$ and $ \{Z(s),s\in S\}$ a latent spatial  process satisfying the equation \ref{eqLienXandZ}. Given a return period $m$ and threshold $u<x_{m}$, the return level $x_m$ for all $s\in S$ is a solution of the following two equations:
\begin{enumerate}
\item[i)] Return period as expected number of events
\begin{eqnarray} \label{eqNewENE}
1=\displaystyle\sum_{t=t_{0}}^{t_{0}+n_{x}m}\left\lbrace c_{\theta}\left(\frac{t}{n},s\right)\phi_{u}(s)\bar{F}_{Z,s}(x_{m}-u)\right\rbrace, 
\end{eqnarray}
\item[ii)] Return period as expected waiting time
\begin{eqnarray}\label{eqNewEWT}
n_{x}m=1+\displaystyle \sum_{i=1}^{\infty}\prod_{t=1}^{i}\left\lbrace 1-c_{\theta}(\dfrac{t}{n},s)\phi_{u}(s)\bar{F}_{Z,s}(x_{m}-u)\right\rbrace,
\end{eqnarray}
\end{enumerate}
where $\bar{F}_{Z, s}$ is  a  survival of generalized Pareto distribution  of $\ell$-exceedances at position $s$; $ \phi_{u}(s)=P(Z(s)>u)$ is the probability of exceedances, $c_{\theta}$ is the tail trend function and $ n_{x}$ is the number of days in the year.
\end{proposition}
The return levels derived from the equations (\ref{eqNewENE} \& \ref{eqNewEWT}) are evaluated by numerical algorithms taking into account the information from the extrapolation of the trend function $c_{\theta}$ on the one hand and the spatial dependence structure of the latent stationary process $Z$ on the other hand. To derive the return  level at points where we have not observations, we assume, the spatial process $Z$ is locally stationary in space.  Under these new assumptions and using the knowledge of the spatial dependence structure of the latent process $Z$ and the equation(\ref{eqzobs}),  we propose the following result: 

\begin{proposition}\label{propRLNonStatRL}
Let $\{X_{t}(s),s\in S, t\in T\}$ be a non-stationary spatio-temporal stochastic process defined on a region $S\subset\mathbb{R}^{d}$ and $ \{Z(s),s\in S\}$ a  latent spatial process. The non-stationary return level $x_{m}(s), s\in S$ of the non-stationary process $X$  is deduced from the return level $z_{m}$ of the latent spatial  process $Z$  such that:
\begin{eqnarray}\label{eqSpatialRL}
x_{m}(s)=z_{m}(s)\widehat{c}_{\theta}(t_{m},s)^{\widehat{\gamma}}+\dfrac{\widehat{c}_{\theta}(t_{m},s)^{\widehat{\gamma}}-1}{\widehat{\gamma}}\left( \widehat{\tilde{a}}_{n}-\widehat{\gamma} \widehat{\tilde{b}}_{n}\right),~~ s\in S,
\end{eqnarray}
where $\widehat{\tilde{a}}_{n}$, $\widehat{\tilde{b}}_{n}$ $\widehat{c}_{\theta}$ and $\widehat{\gamma}$ are the respective estimators of  $\tilde{a}_{n}$, $\tilde{b}_{n}$, $c_{\theta}$ and $\gamma$. $t_{m}=1+\dfrac{n_{x}m}{n}$ with $n_{x}$ the number of days in the year and n the size of the sample observed.
\end{proposition}
This result is a consequence of the equation (\ref{eqzobs}) and will be used to derive the non-stationary return levels of process $X$, using the $z_{m}$ computed from the estimated parameters. Thus, to calculate the  $z_{m}$  at grid points where we have not observations, we use a spatial model from marginal parameters to build a spatial map on the domain of study based on covariates such as longitude and latitude. 

\section{Application to extreme precipitation in Burkina Faso}\label{secPrecipBF}

\subsection{Data set analysis}
\begin{figure}[ht!!!!!!!!!!!]
\centering
\includegraphics[scale=0.35]{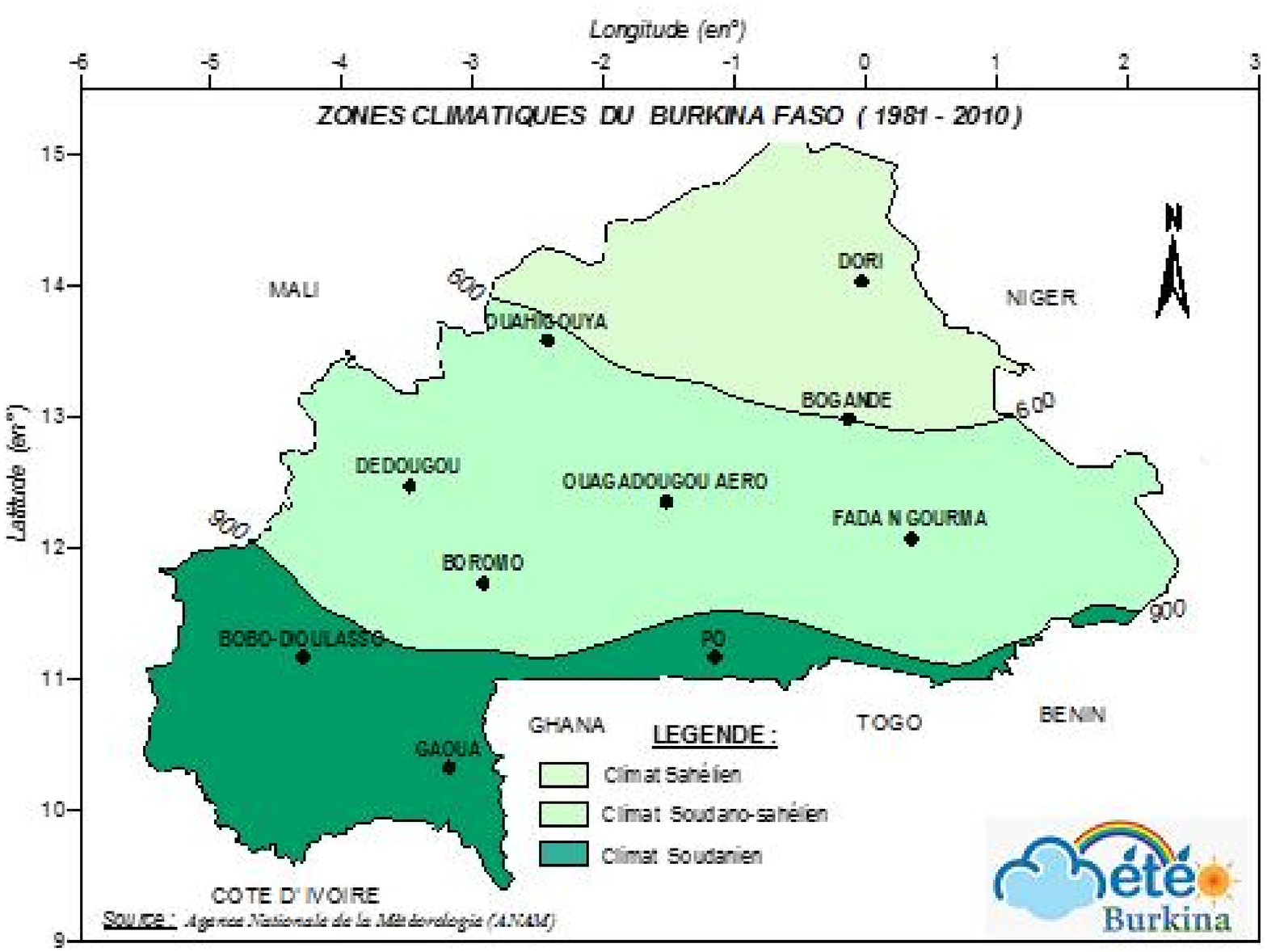}
\caption{Spatial distribution of synoptic sampling stations associated with climatic regimes, source: National Meteorology Agency of Burkina Faso.}\label{figstations}
\end{figure}

This study uses time series of daily precipitation measurements from 1957 to 2016 provided by ten synoptic stations extracted from the Burkina Faso climatological database.  These stations have been selected to ensure good spatial uniformity and representativeness of different climatic regimes and data quality. The figure\ref{figstations} gives the spatial distribution of the synoptic stations in our study area.  In order to limit the problems related to seasonal rainfall cycles on each station,  we worked from the sub-series corresponding to rainy days. The period from may to october was chosen because  during this period that the most rainfall is recorded in Burkina Faso.
Thus, a serie of 60 by  184 days is extracted to constitute the time series of daily rainfall. On these time series, we apply a run declustering procedure with a daily step (r=1 day) to identify the groups of approximately independent extreme observations within the sample in order to avoid short-term dependencies in the time series.

\subsection{Marginal characteristics}
\begin{figure}[ht!!!!!!!!!!]
\centering
\includegraphics[scale=0.37]{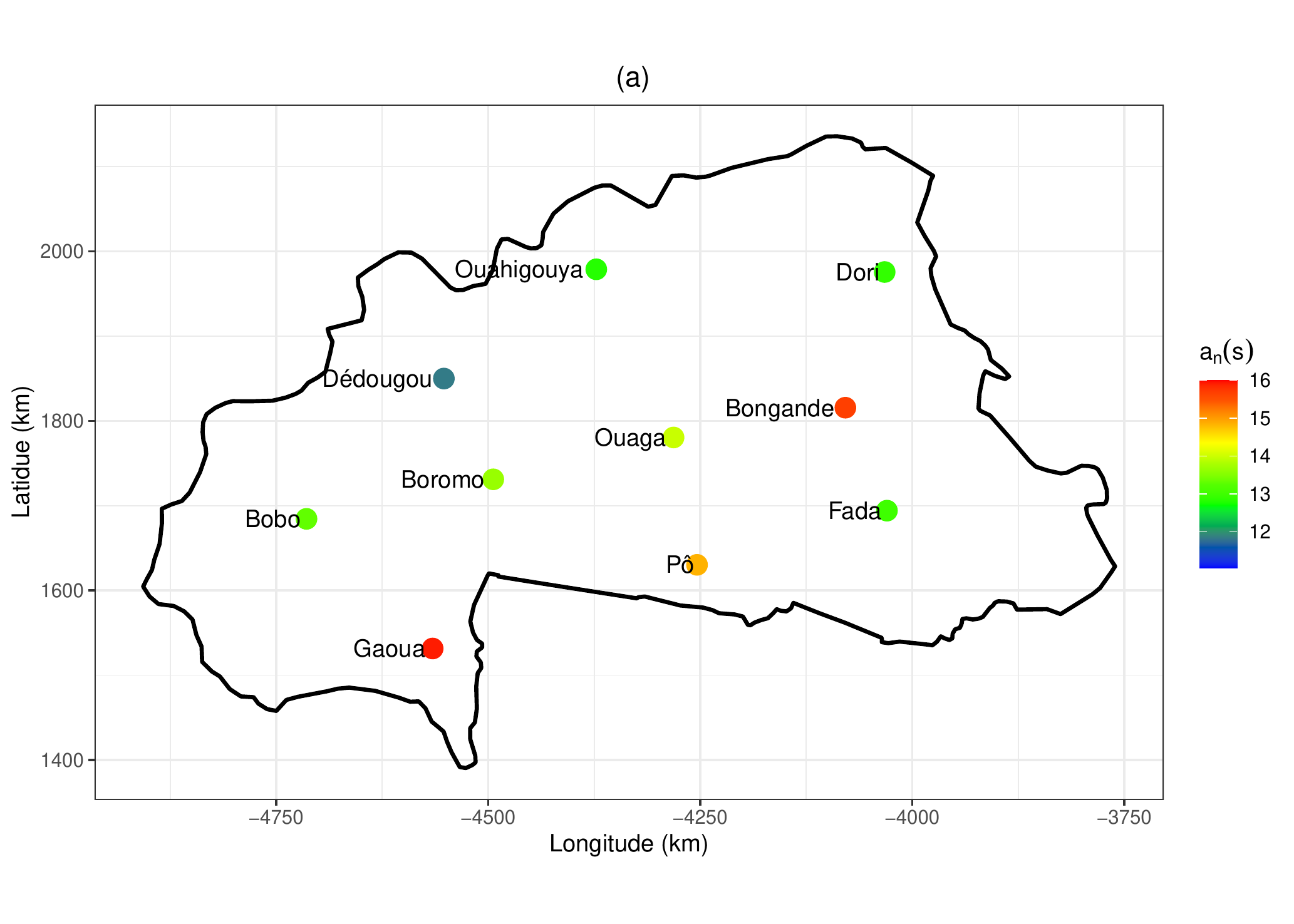}
\includegraphics[scale=0.37]{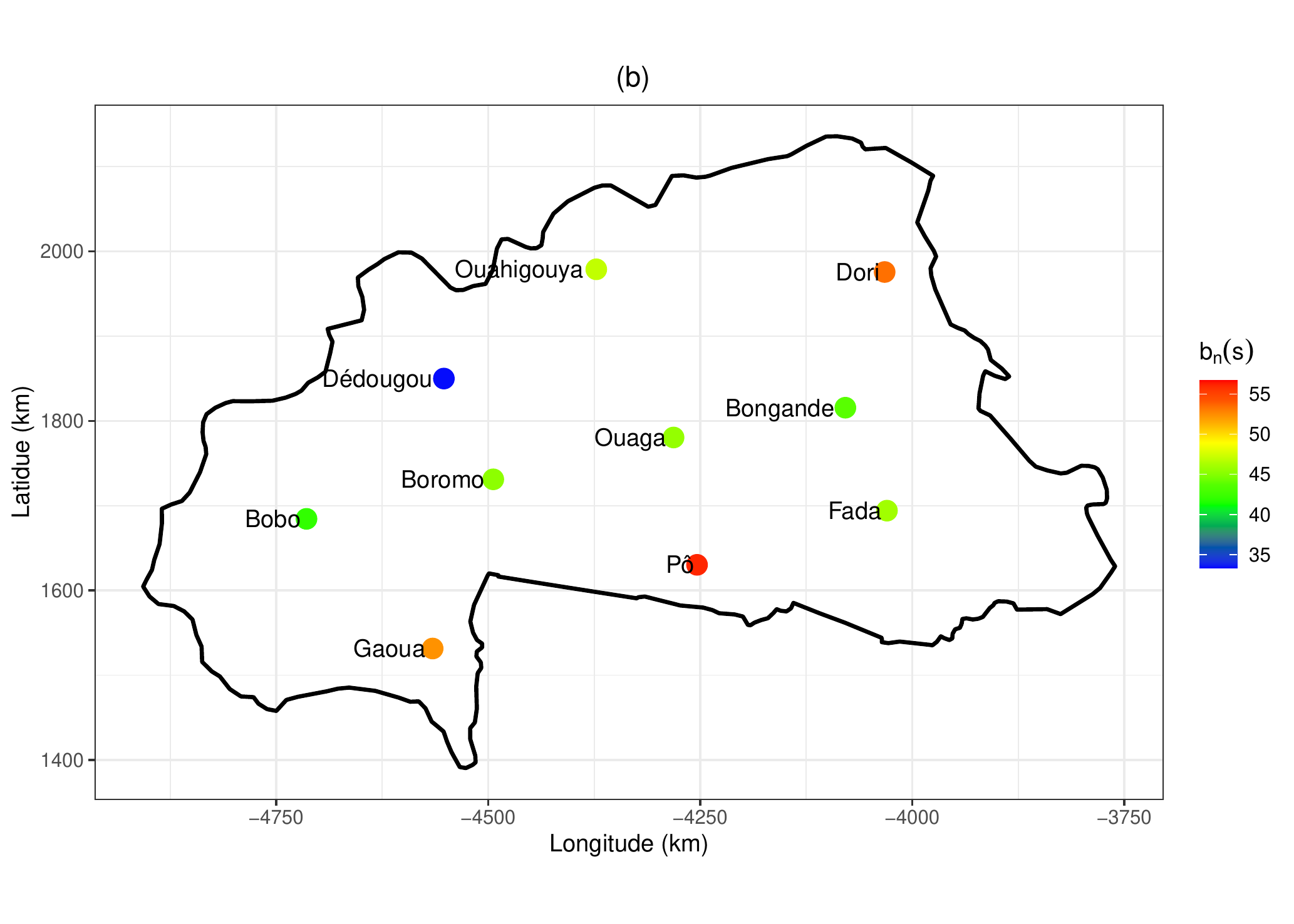}
\includegraphics[scale=0.37]{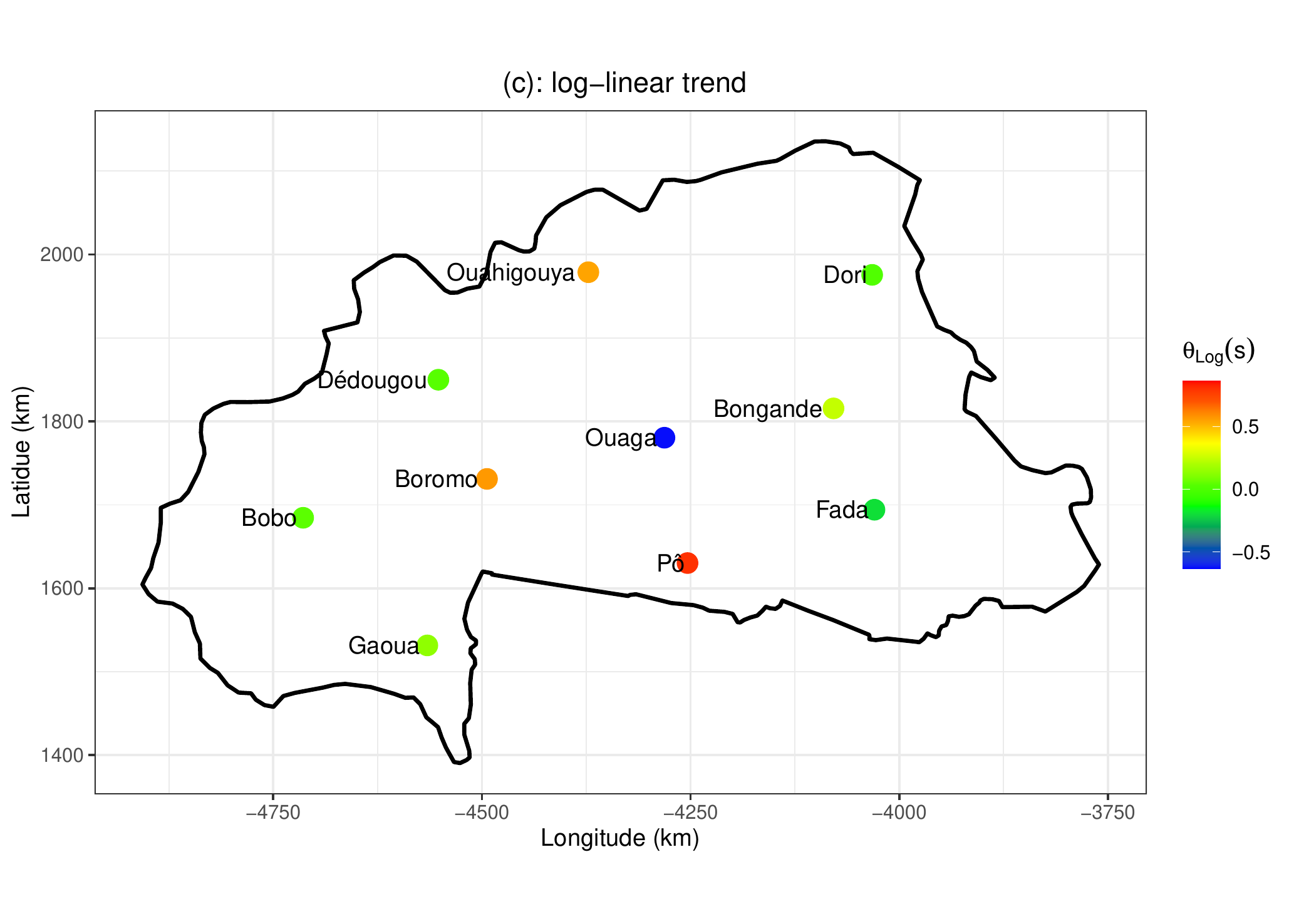}
\includegraphics[scale=0.37]{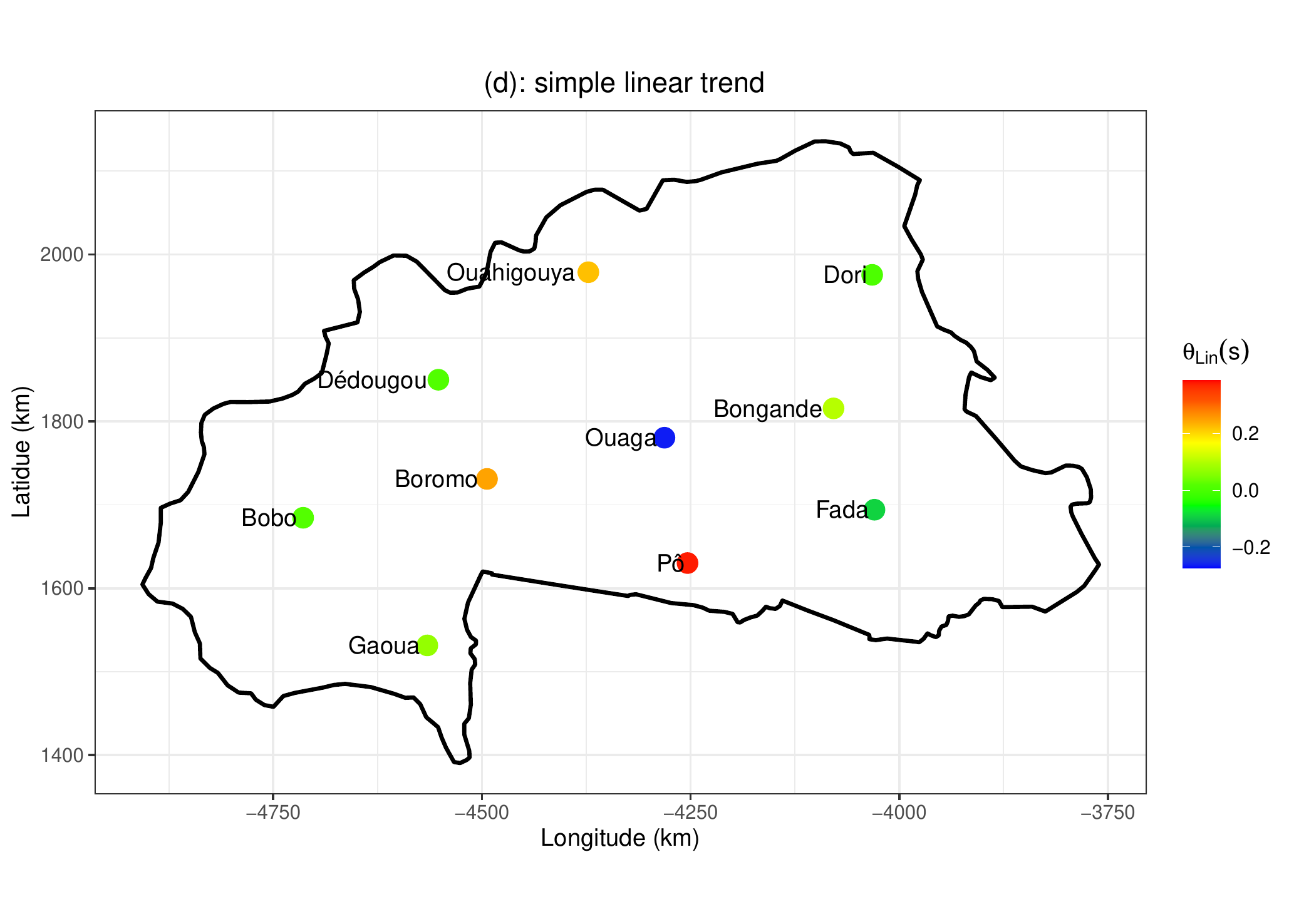}
\caption{Estimated functions $a_{n}$ (\ref{fig_parsMarginal}a) and $b_{n}$ (\ref{fig_parsMarginal}b) of the generalized $\ell$-Pareto process for modelling extreme precipitation in Burkina Faso, and $\theta$ for Log-linear (\ref{fig_parsMarginal}c) and simple linear(\ref{fig_parsMarginal}d) trends. Estimates are obtained by shifting the local empirical quantiles $u_{0.95}\left\lbrace Z(s)\right\rbrace $ by $\widehat{\tilde{b}}_{n} =56.1 mm$.}\label{fig_parsMarginal}
\end{figure}

\begin{figure}[ht!!!!!!!!!!!!!!!!!!!!]
\centering
\begin{tabular}{rrr}
 Sudanian climate ~~~~~~~  & Sudano-Sahelian climate  & Sahelian climate  \\ 
  \hline
  %Bobo ~~~~~~~~~~~ &Ouaga~~~~~~~~~~~~~&Dori~~~~~~~~~~~~~\\
 \includegraphics[scale=0.22]{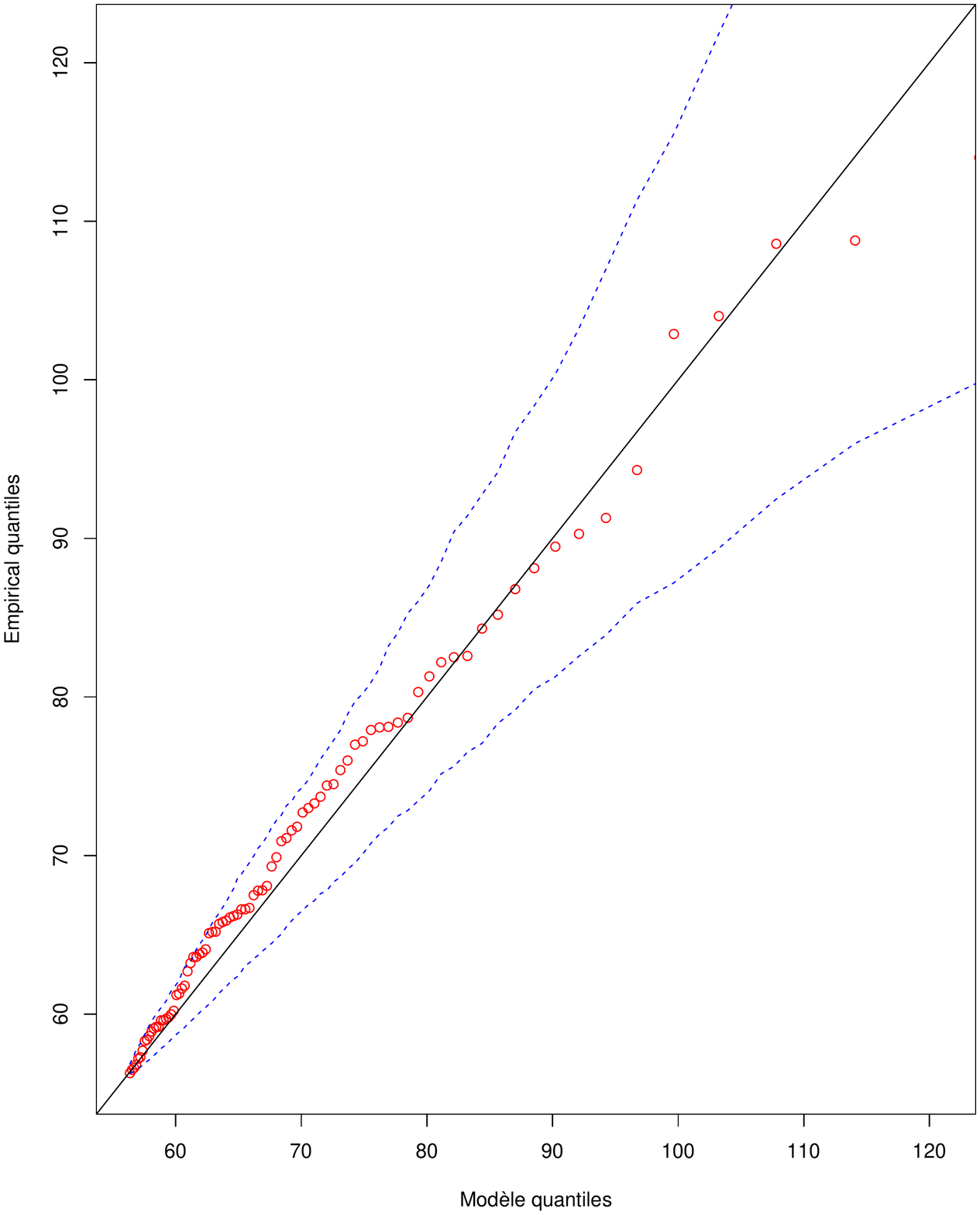}& \includegraphics[scale=0.22]{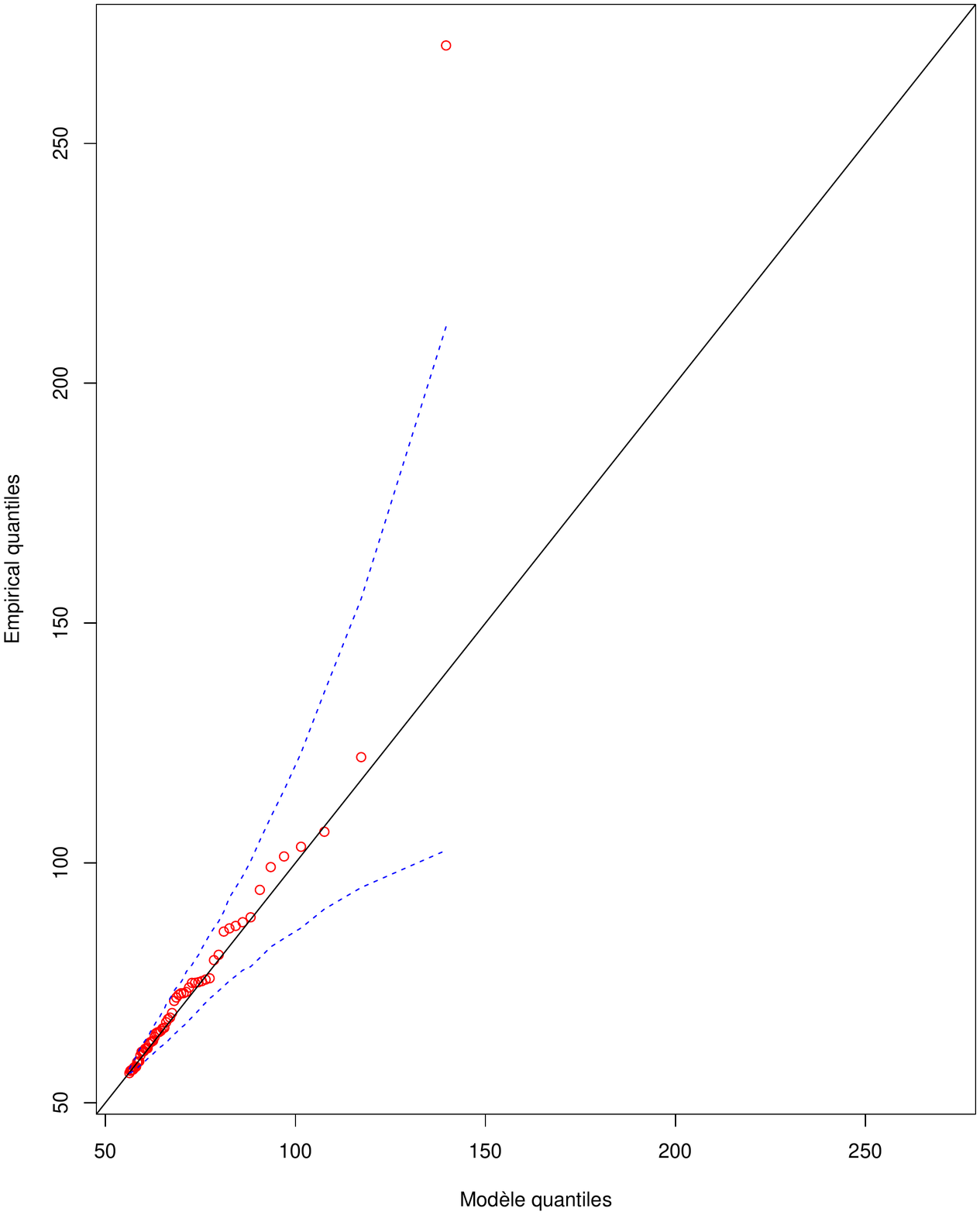}& \includegraphics[scale=0.22]{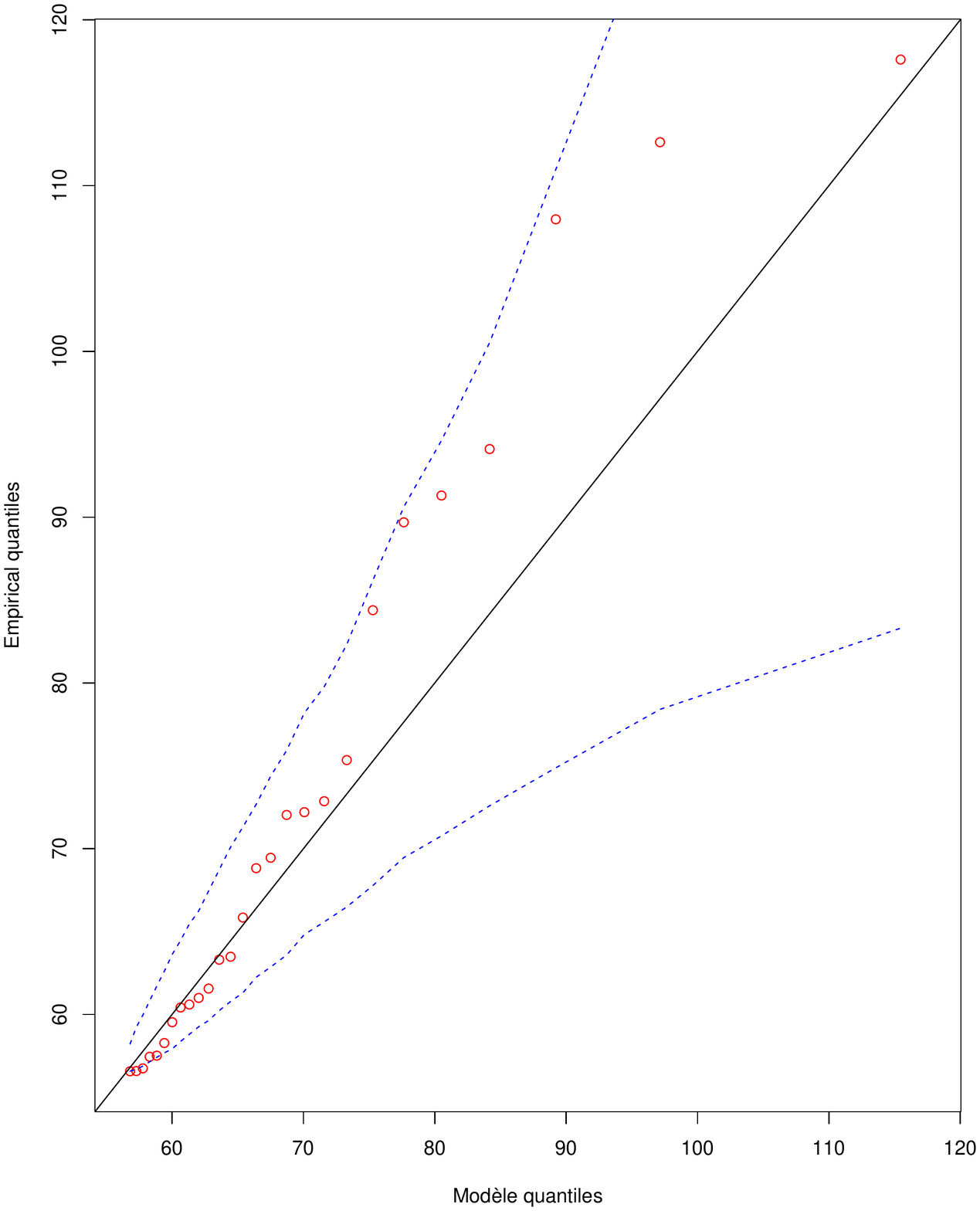}\\
 % Pô~~~~~~~~~~~~~~ & Boromo~~~~~~~~~~~~~& Ouahigouya~~~~~~~~\\
\includegraphics[scale=0.22]{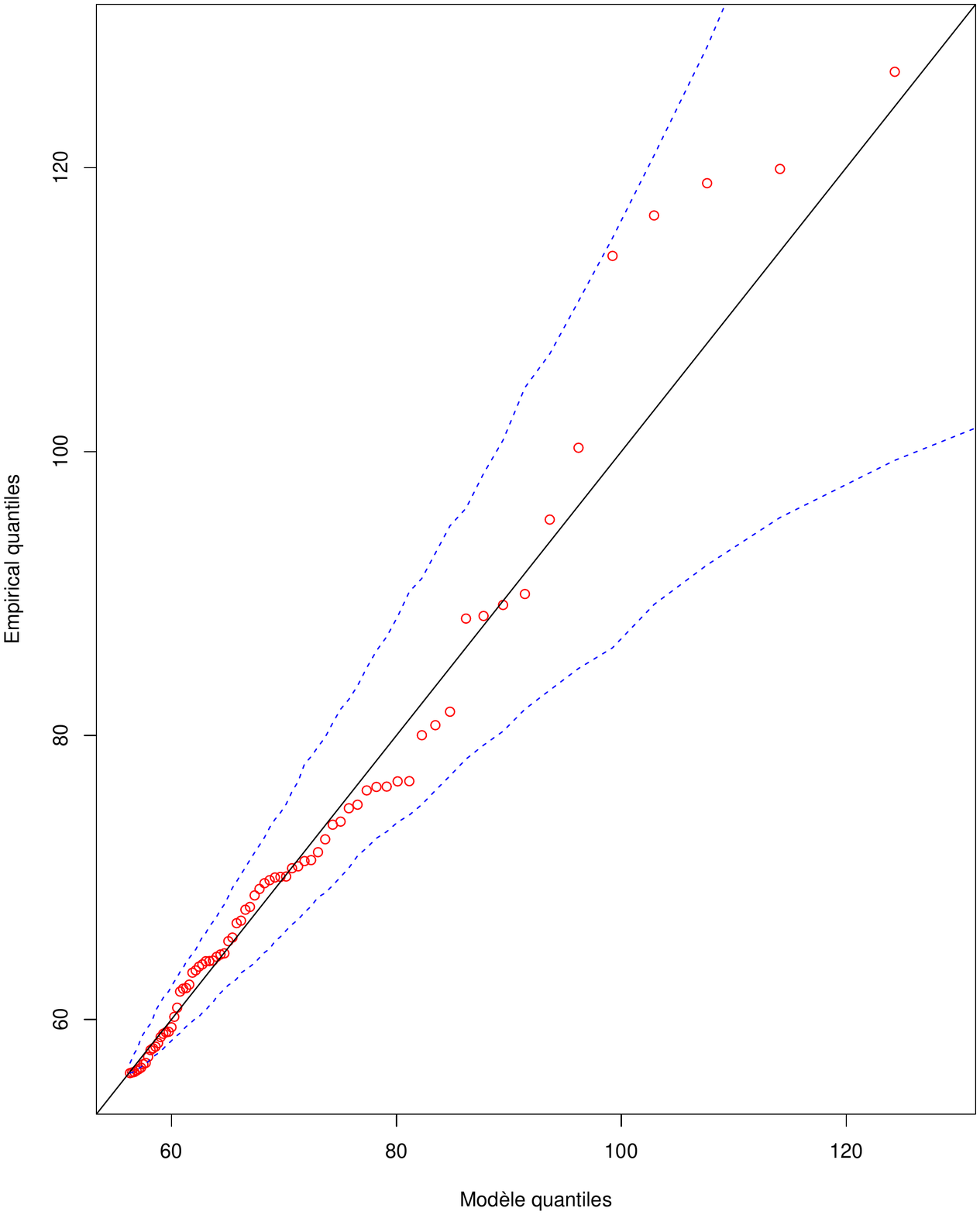} & \includegraphics[scale=0.22]{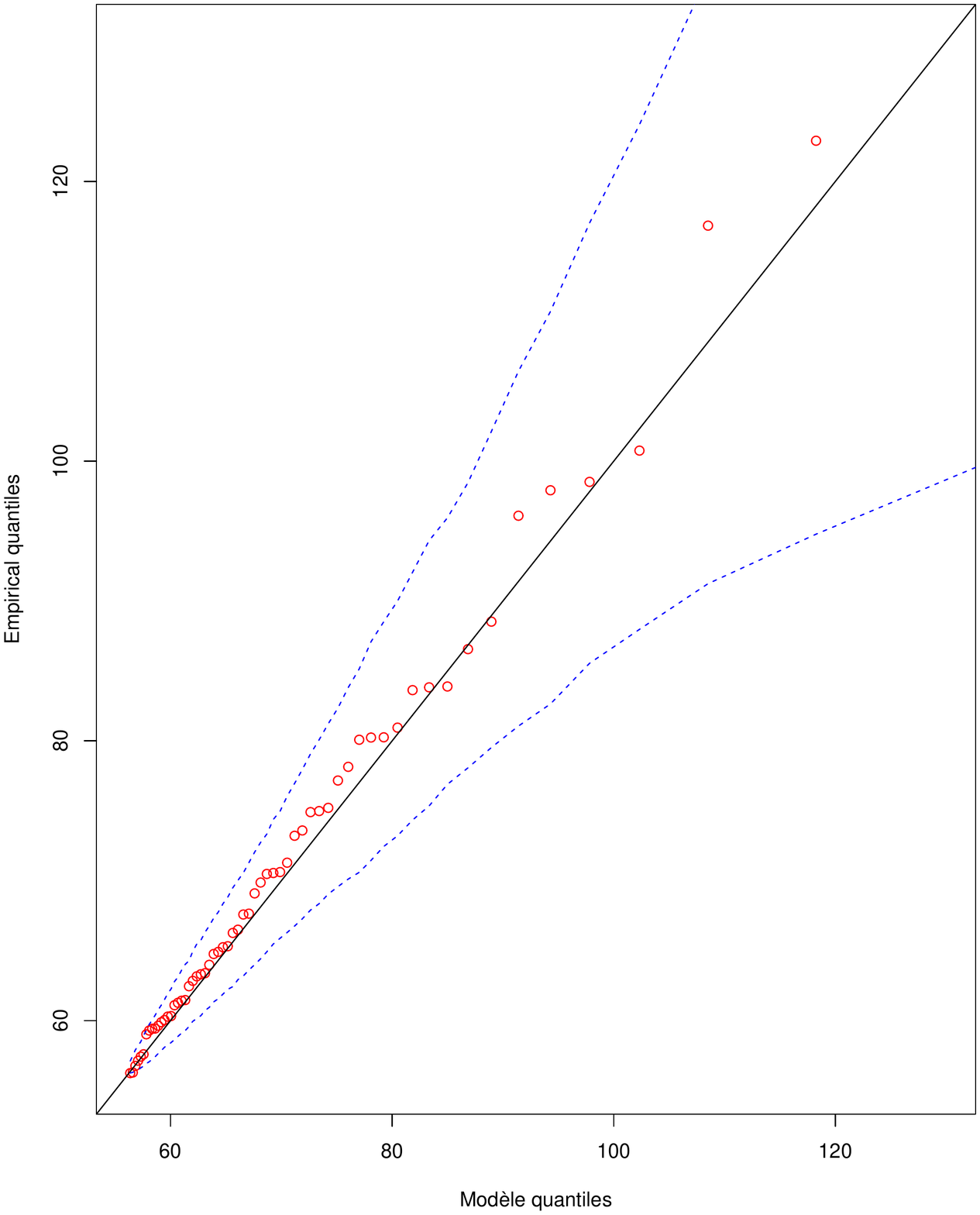}&\includegraphics[scale=0.22]{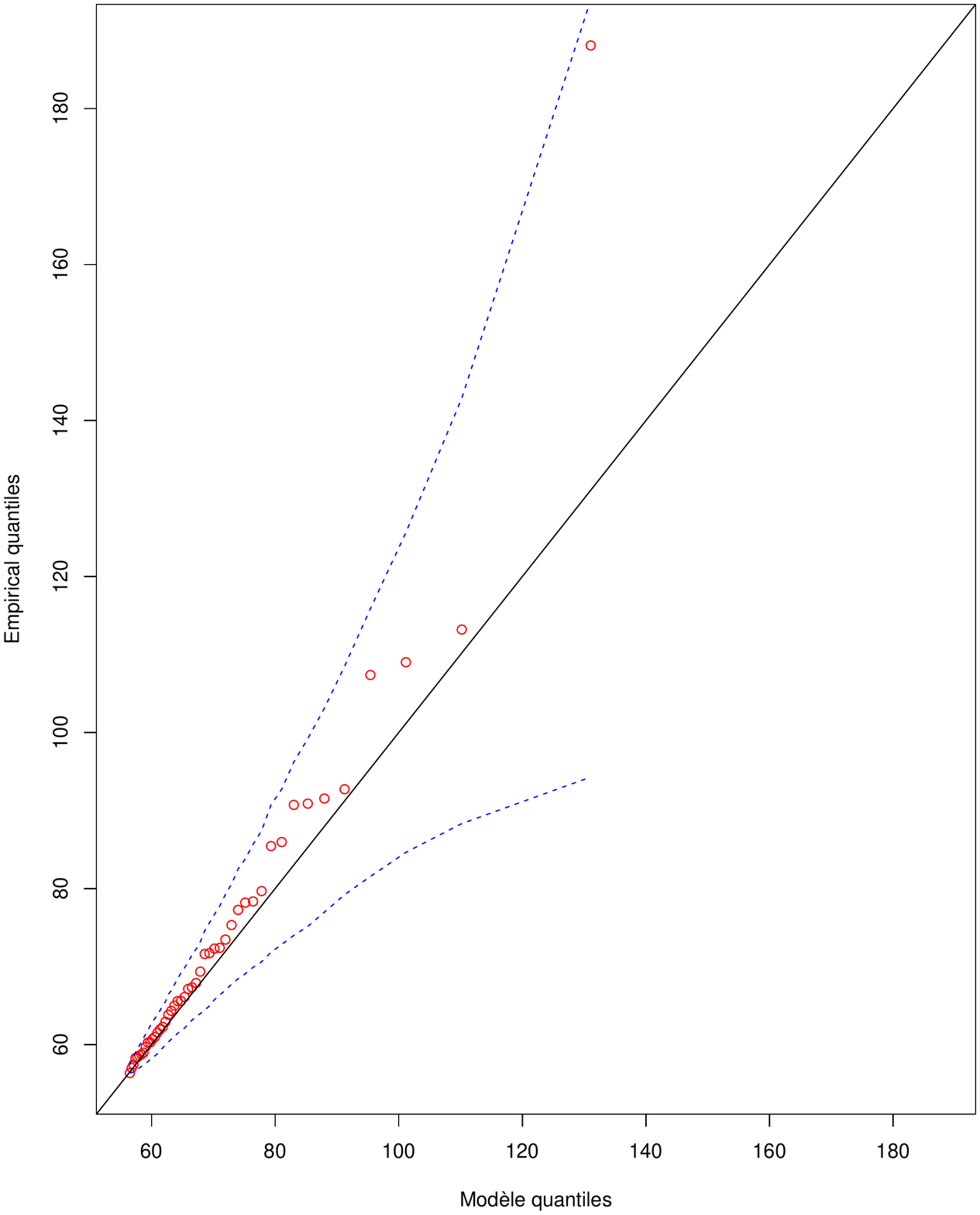}\\
  \hline
\end{tabular}
\caption{QQplot of local tail distributions of six synoptic positions. The thresholds correspond to the local 0.865 quantiles of the $\ell$-exceedances.}\label{figQQplots}
\end{figure}

\begin{figure}[ht!!!!!!!]
\centering
\includegraphics[scale=0.75]{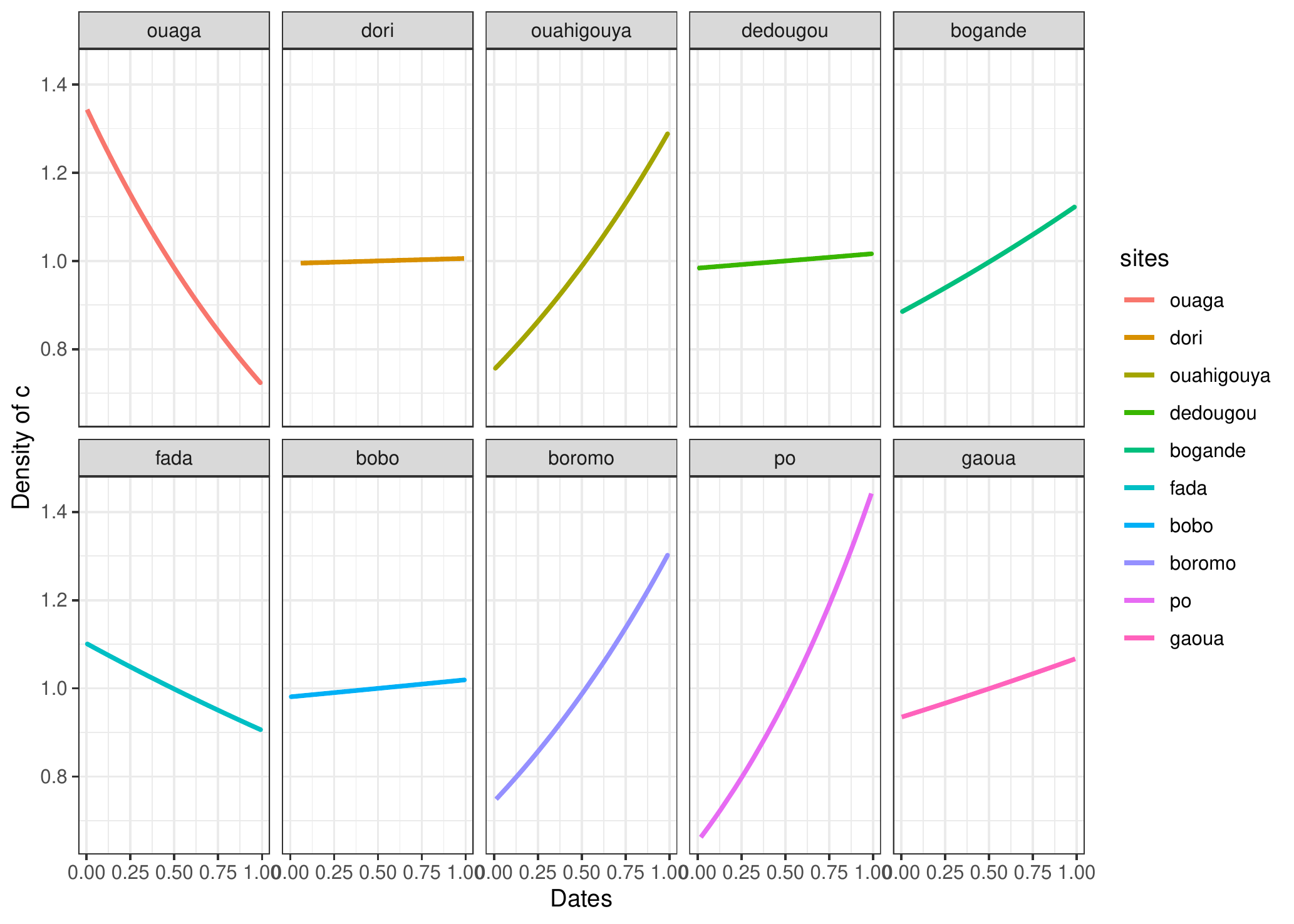}
\caption{Local adjustment of evolution of the frequencies of extreme precipitation by a log-linear model trend  $c_{\theta}^{1}$.}\label{figTrendsFunction}
\end{figure}

The scale $a_{n}(s)$,  location $b_{n}(s)$ and parameters of log-linear and linear trend function $c_{\theta}$ are estimated for any $s\in S$ and are shown respectively on maps (a), (b), (c) and (d) of figure(\ref{fig_parsMarginal}).
In practice we deduce from the generalized $\ell$-Pareto model fitted to the data, the marginal tail behavior of the survival distribution $\bar{F}_{Z,\widehat{\gamma}(s),\widehat{\sigma}(s)}$ for any $s\in S$ from the equation (\ref{eqLparetoMargins}) given a sufficiently large $u_{q} >0$ threshold.
\begin{eqnarray}\label{eqLparetoMargins}
P\left[Z(s)-u_{q}(s) \geq z \mid Z(s)\geq u_{q}(s)\right]\approx  \bar{F}_{Z,\widehat{\gamma}(s),\widehat{\sigma}(s)}(z), ~~ z\geq 0, 
\end{eqnarray}
with $\widehat{\sigma}(s)=\widehat{a}_{n}(s)+\widehat{\gamma}\left(u_{q}(s)-\widehat{b}_{n}(s)\right)$, $s\in S$, where $\widehat{a}_{n}$, $\widehat{b}_{n}$, and $\widehat{\gamma}$ are the marginal parameters estimators described in section (\ref{secMarginal}). 
We can use it to check the quality of the marginal adjustment of the stationary process $Z$.  Figure \ref{figQQplots} shows the QQplots of the local tail distribution due to two stations by climatic zone. Columns 1, 2, and 3 of Figure \ref{figQQplots} represent the respective adjustments for the Sudanian, Sudano-Sahelian, and Sahelian zones. Globally, the fit of the marginal models seems convincing, as most of the observations remain within the confidence intervals. 
% latex table generated in R 3.6.3 by xtable 1.8-4 package
% Sun Oct 11 14:29:04 2020
Furthermore, the sign of the $\theta$ parameter of the trend function tells us that the frequencies of extreme precipitation are locally quite variable throughout the study region. They have an increasing trend ($\theta > 0$) in areas such as Ouahigouya, Bogande, Boromo, Gaoua, and Po. On the other hand, at the Ouagadougou and Fada stations, extreme rainfall frequencies tend to decrease ($\theta<0$).  Figure \ref{figTrendsFunction} gives us the details of the adjustment of the tail trend function by a log-linear model.  
\subsection{Estimated spatial dependence model}\label{secDependenceEstimate}
We use a spatial model of the marginal parameters dependent on significant covariates  to obtain the marginal parameters at the different grid points using a generalized linear regression model, in order to prepare the ground for calculations of non-stationary return levels at locations where we have no observations.
\begin{eqnarray}\label{eqSpatialMarginModel}
\left\{\begin{array}{rl}
 a_{n}(s)  =  a_{0} + a_{1}Long\times Lat(s) ~~~~~~~~~~~~~~~\\
 b_{n}(s) =  b_{0} +b_{1}Lat^{2}(s)+b_{2}Long\times Lat(s)\\
 \theta(s)=\theta_{0}+ \theta_{1}\times Long +\theta_{2}Lat~~~~~~~~~~~~~~\\
\gamma(s)=\gamma_{0} ~~~~~~~~~~~~~~~~~~~~~~~~~~~~~~~~~~~~~~~~~~

\end{array} 
\right., ~~s\in \mathcal{N}_{s_{0}},
\end{eqnarray}
where $\mathcal{N}_{s_{0}}$ is a small neighborhood around $s_{0}$. In what follows, these regional neighborhoods will be determined by a small number, $D_{0}$ of nearest stations from the site $s_{0}$,  thus we will write $\mathcal{N}_{s_{0}}=\mathcal{N}_{s_{0}, D_{0}}$ . Obviously, the choice of neighbourhood is important; the assumed stationary marginal parameters could be a poor approximation for large neighbourhoods (i.e., for large $D_{0}$ ), while the simulation of the process could be cumbersome for small neighbourhoods characterized by a small $D_{0}$ number of stations. In principle, the choice of $D_{0}$ should be such that the spatial dependence structure and marginal parameters  is approximately stationary within each selected neighborhood $\mathcal{N}_{s_{0}, D_{0}}$ around $s_{0}\in \mathcal{G}$.  We obtain the relatively homogeneous, non-overlapping sub-regions using the k-means clustering method centered on the reference stations. This method is extensively used because it is computationally simple and produces accurate results, compared to other more complex clustering methods. The longitude and latitude of the grid points were used as input variables in the k-means clustering algorithm to form the ten clusters centered on the reference stations. \vspace{0.2cm}

\begin{figure}[ht!!!!!!!]
\centering
\includegraphics[scale=0.37]{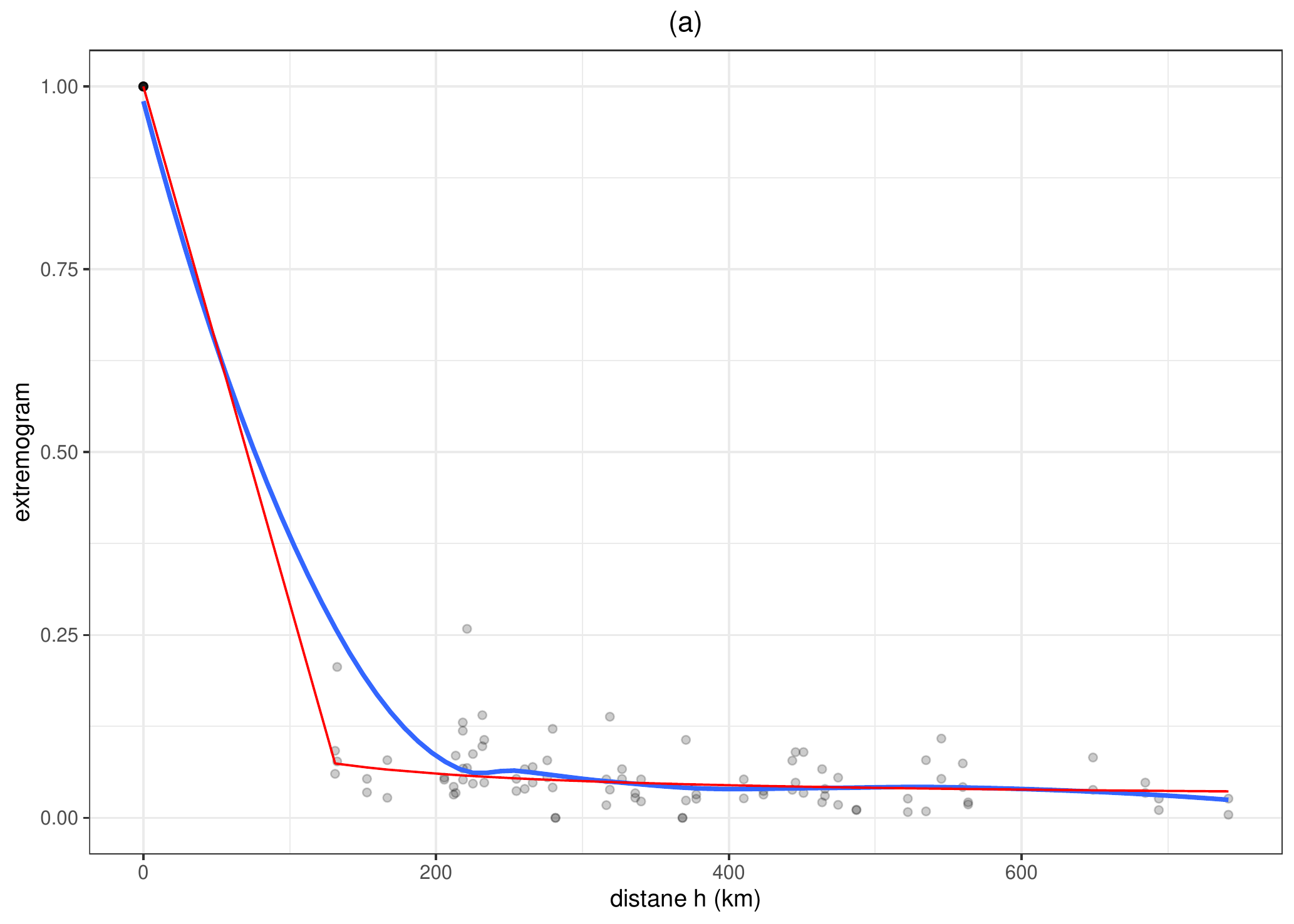}
\includegraphics[scale=0.37]{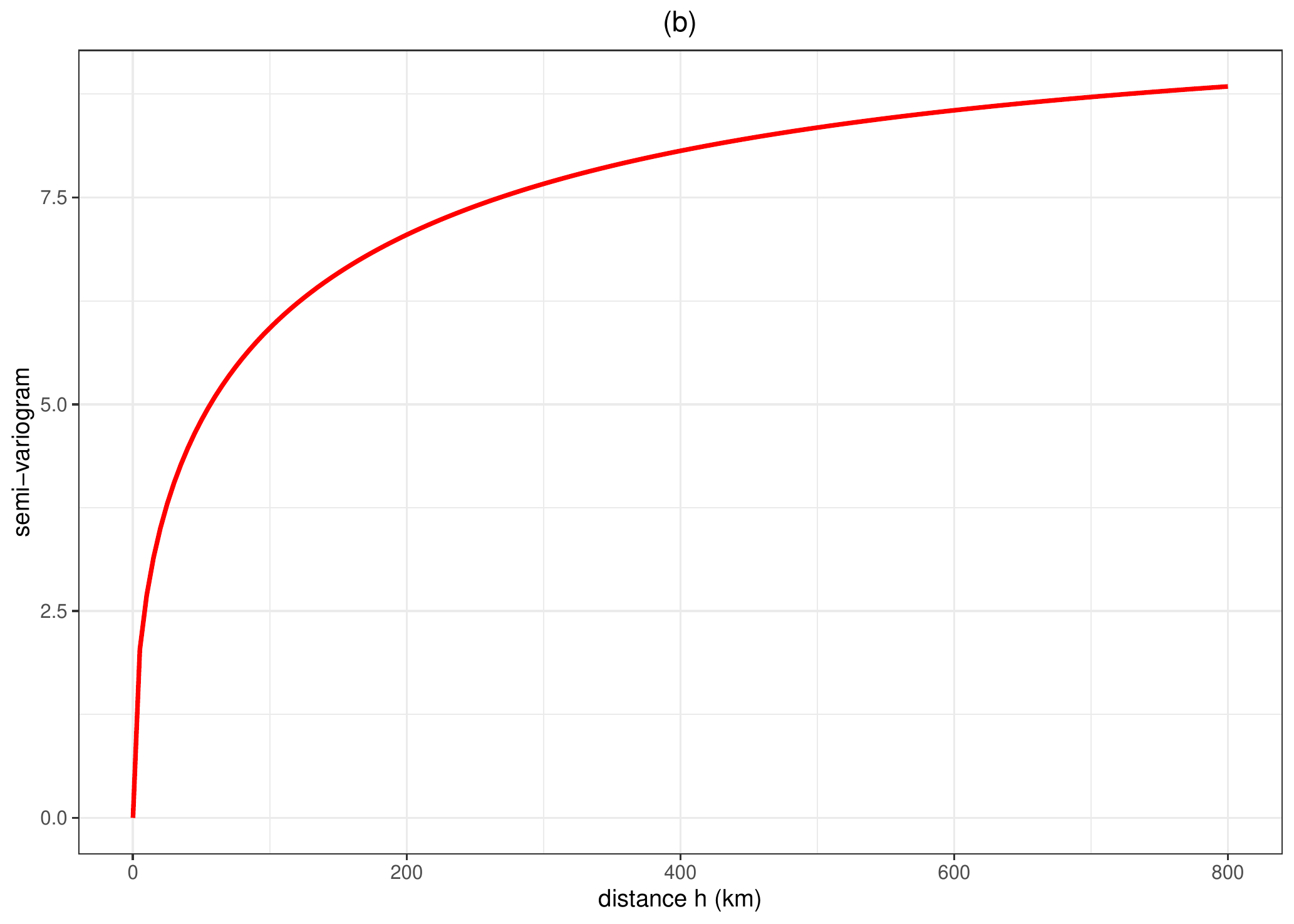}
\caption{Isotropic  extremogram $\pi(h)=\ds\lim_{q\rightarrow 1}P(Z(s+h)>u_{q}(s+h)\mid \{Z(s)>u_{q}(s),\ell(Z)>u\}$ (Figure \ref{figdependence}a) and variogram  (Figure \ref{figdependence}b) estimated for risk functional $\ell(Z)=\max_{s\in S}Z(s)$ of distance between locations $s$ and $s+h$.}\label{figdependence}
\end{figure}

In addition, to better take into account the dependence structure of precipitation data and capture the possible isotropic, we choose a flexible parametric semi-variogram $\nu$ belonging to the power class models such that $\nu(h)=\left(\|h\|/\tau\right)^{\kappa}$, with $\tau>0$, $h\in\mathbb{R}^{2}$, $\kappa\in\left] 0,2\right]$.
We check the adequacy of the fit of the model to the data using the extremogram and the variogram. Figures (\ref{figdependence}a) and (\ref{figdependence}b) show respectively the good quality of the fit of the dependence measures such as the semi-variogram and  extremogram. In Figure(\ref{figdependence}a), the points in gray represent the calculated empirical extremogram and the blue curve is the fitted empirical extremogram, while the red curve represents our fitted dependence model. The red curve in Figure 5b is the variogram model fitted to the data as a function of distance. 
It is noted that extreme precipitation is spatially correlated for distances of the order of 200km.  This spatial dependence decreases gradually as the distance increases before stabilizing for example for a value of $\pi(h)=0.0625$. This reflects the very localized nature of extreme rainfall and it comforts us in our analysis because this property is generally present in rainfall data.

\subsection{Non-stationary return level results }
The return levels $x_{50}$ and $x_{100}$ are first estimated punctually (see Figure\ref{figNonStatRLsim})  using the equations (\ref{eqNewENE} \& \ref{eqNewEWT}) before being spatially interpolated to the points where we have no observations using the dependence structure estimated from the observed data and  parameters of trend function spatially extrapolated to the section (\ref{secDependenceEstimate}).
% latex table generated in R 3.6.3 by xtable 1.8-4 package
% Sun Oct 11 14:29:04 2020
\begin{figure}[ht!!!!!!!!!!!!!!!!!!!!]
\centering
\begin{tabular}{rrr}
 Sudanian climate ~~~~~~~  & Sudano-Sahelian climate  & Sahelian climate  \\ 
  \hline
  \\
  %Bobo ~~~~~~~~~~~ &Ouaga~~~~~~~~~~~~~&Dori~~~~~~~~~~~~~\\
 \includegraphics[scale=0.23]{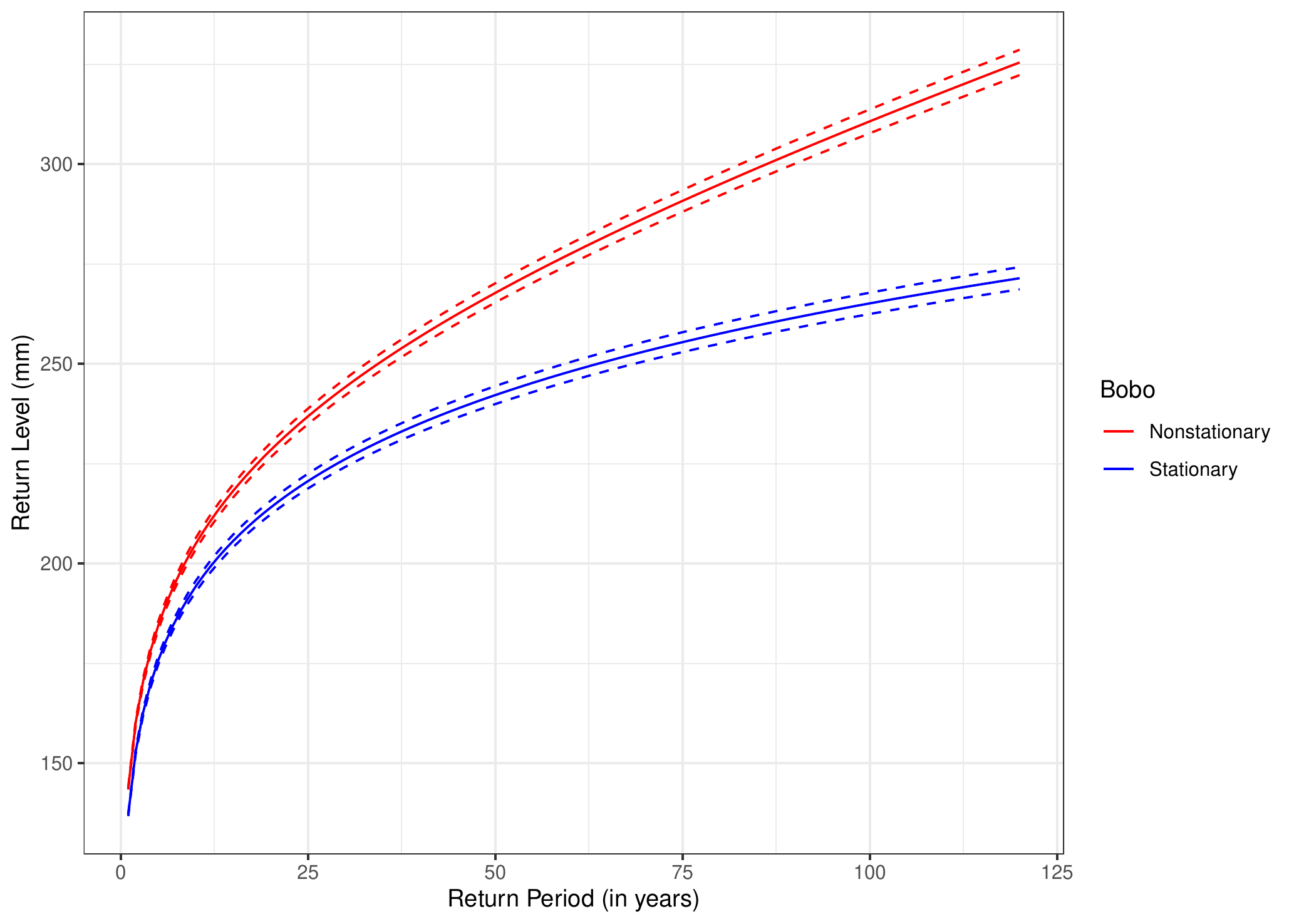}& \includegraphics[scale=0.23]{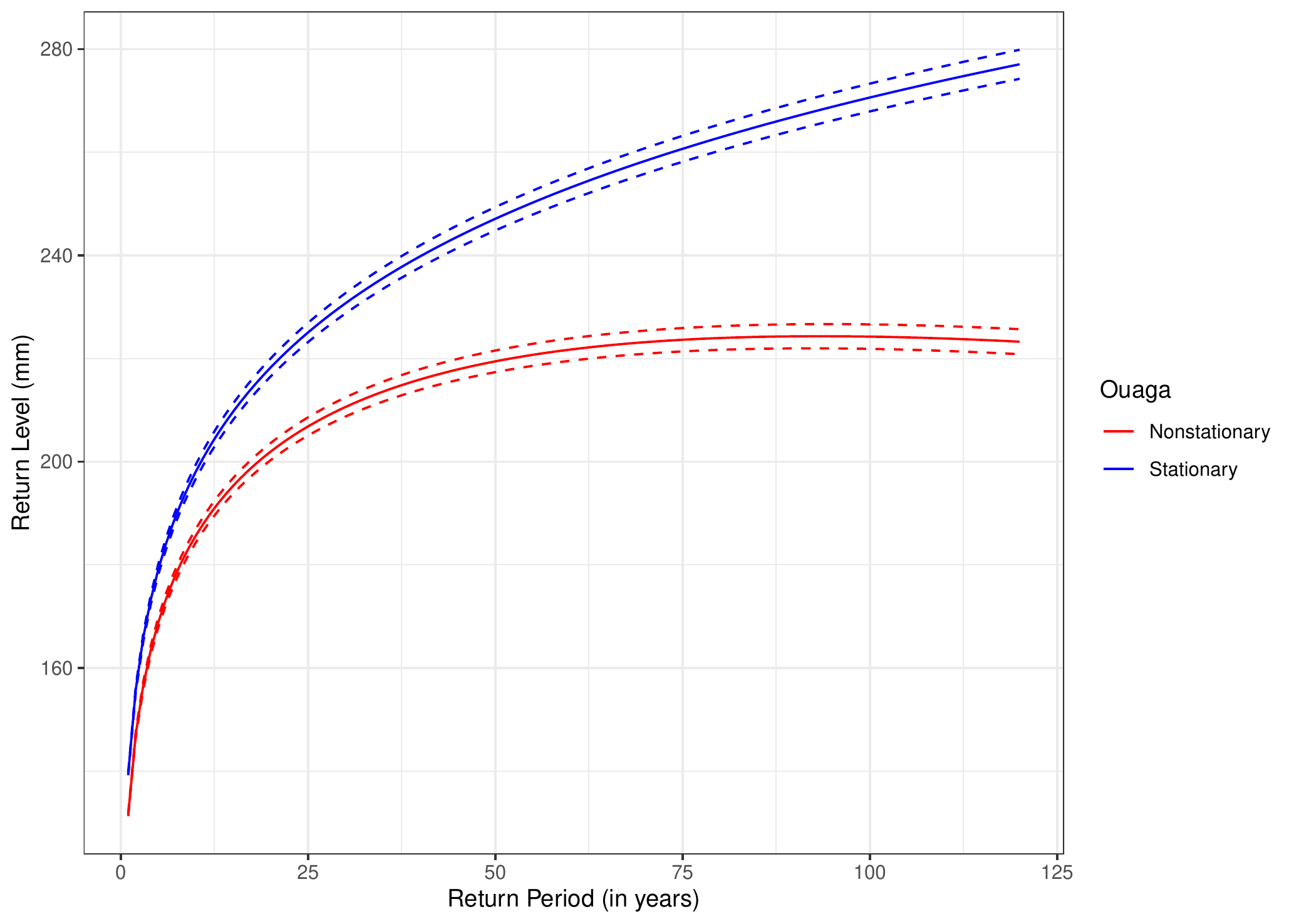}& \includegraphics[scale=0.23]{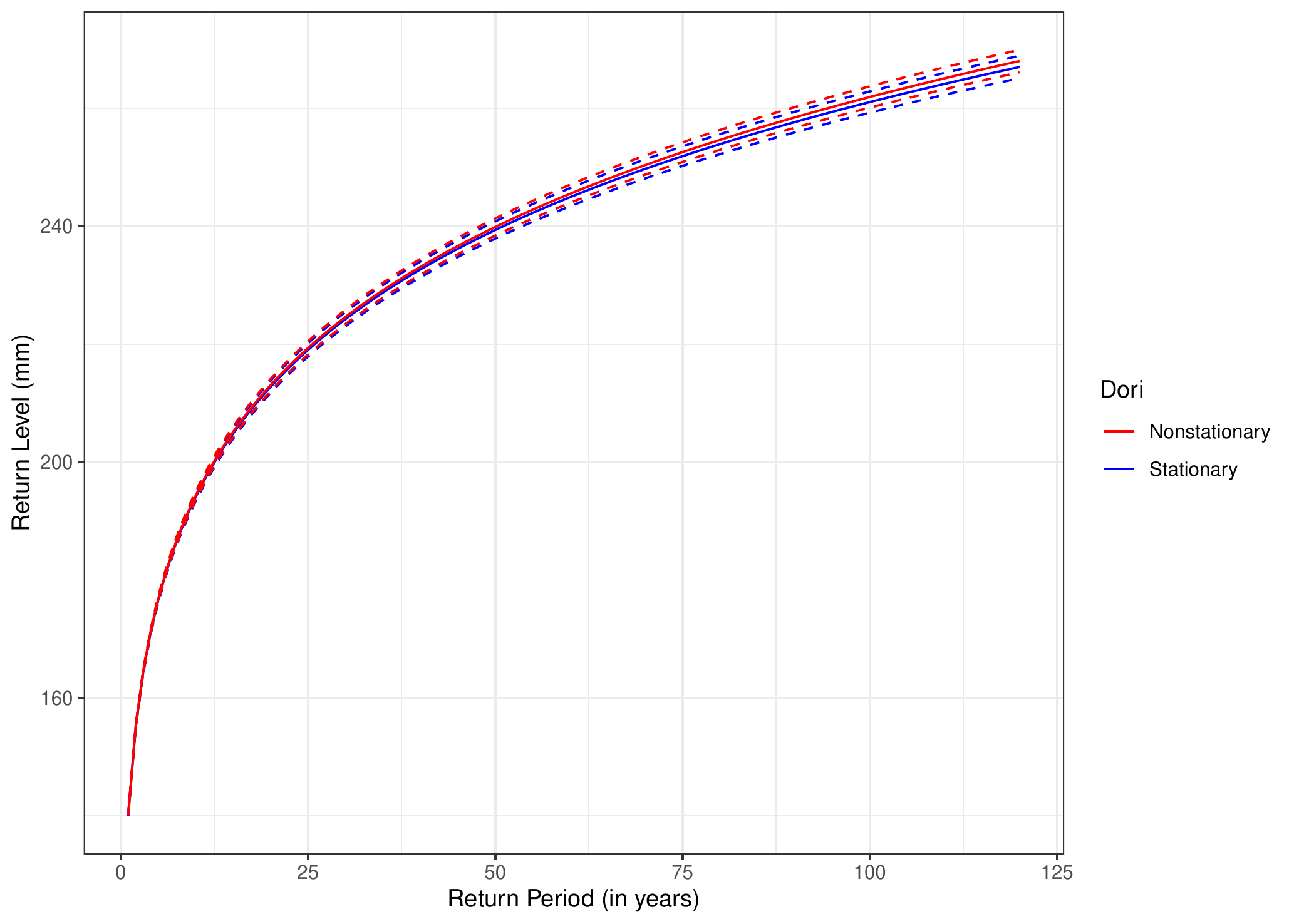}\\
 % Pô~~~~~~~~~~~~~~ & Boromo~~~~~~~~~~~~~& Ouahigouya~~~~~~~~\\
\includegraphics[scale=0.23]{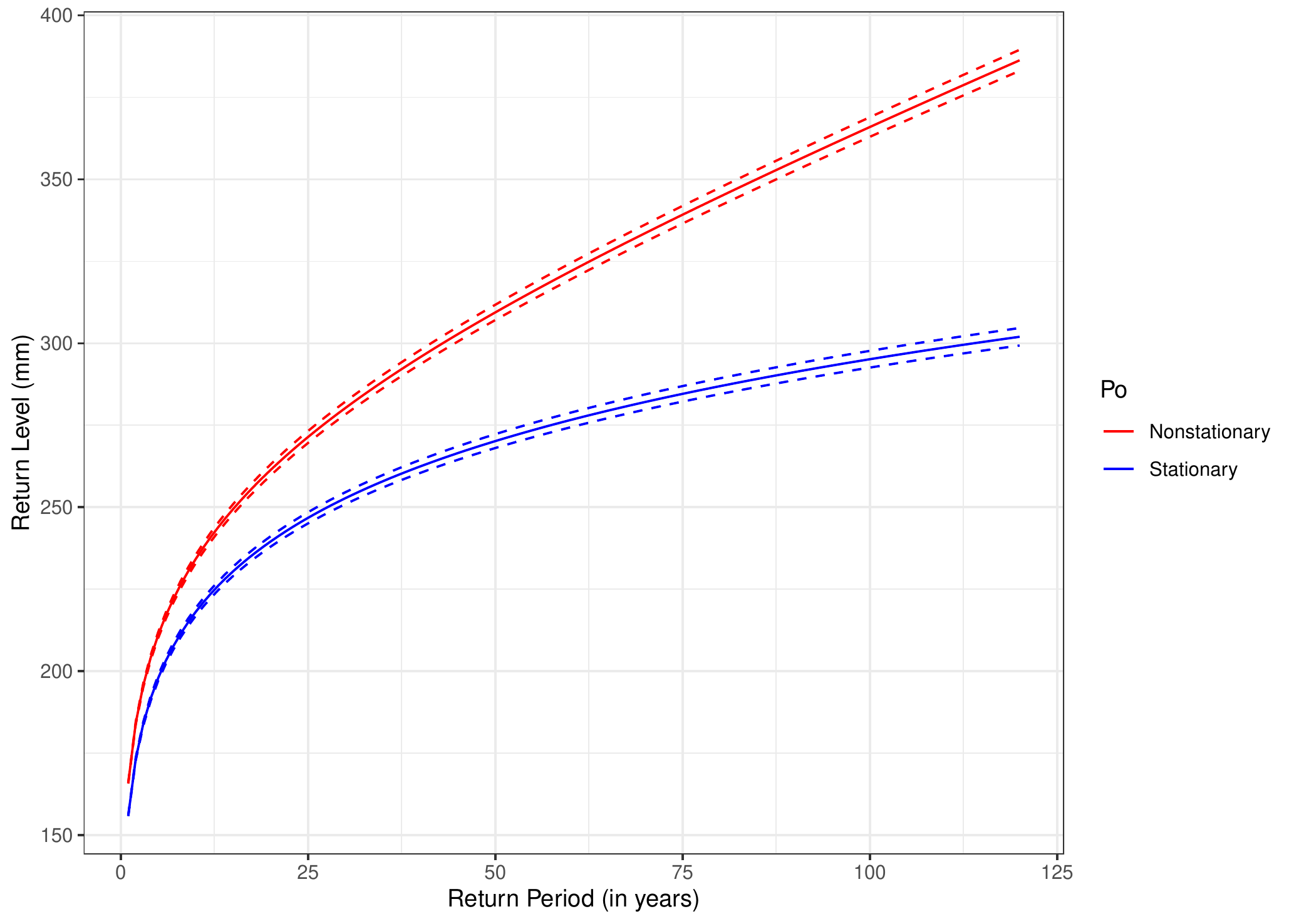} & \includegraphics[scale=0.23]{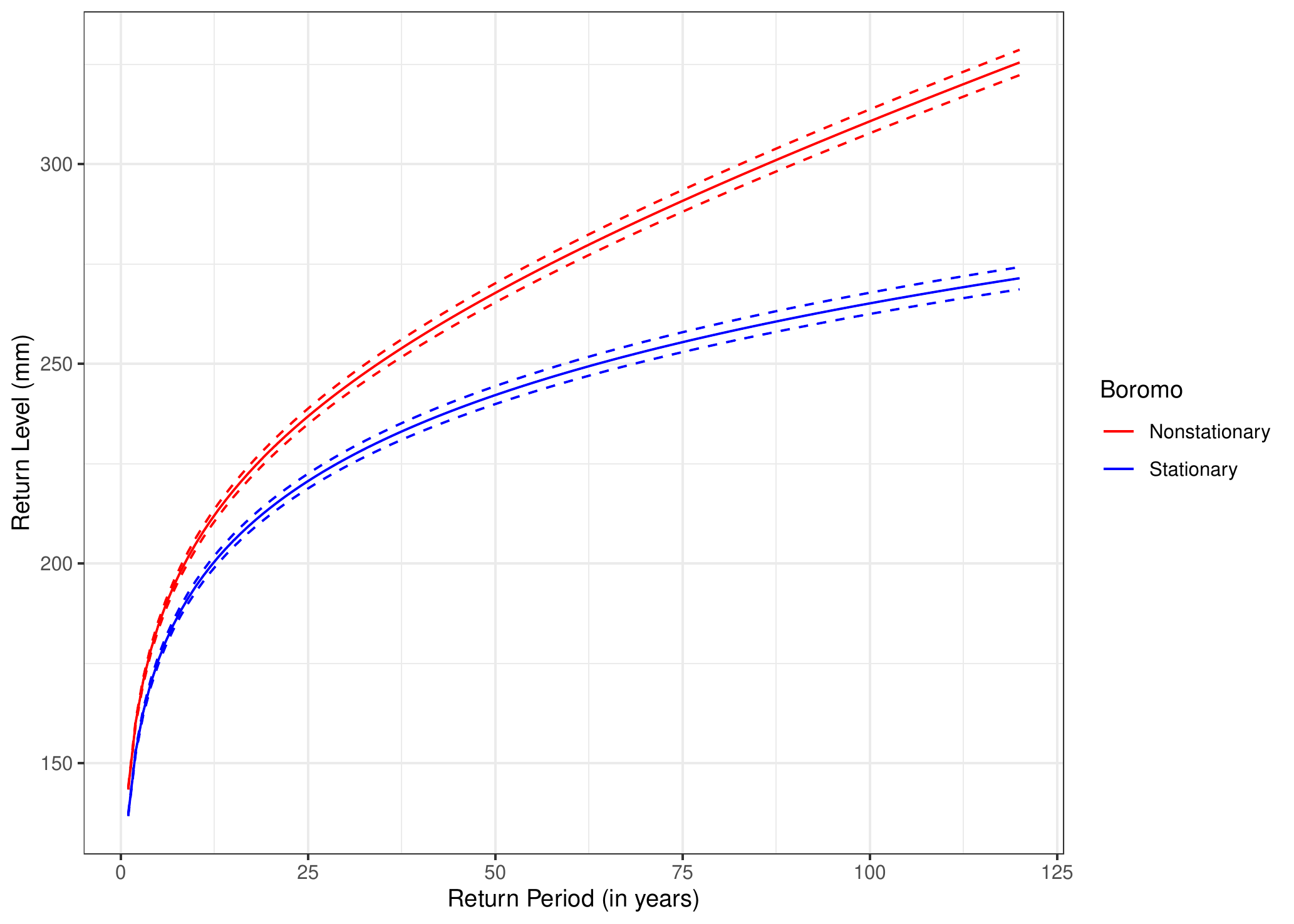}&\includegraphics[scale=0.23]{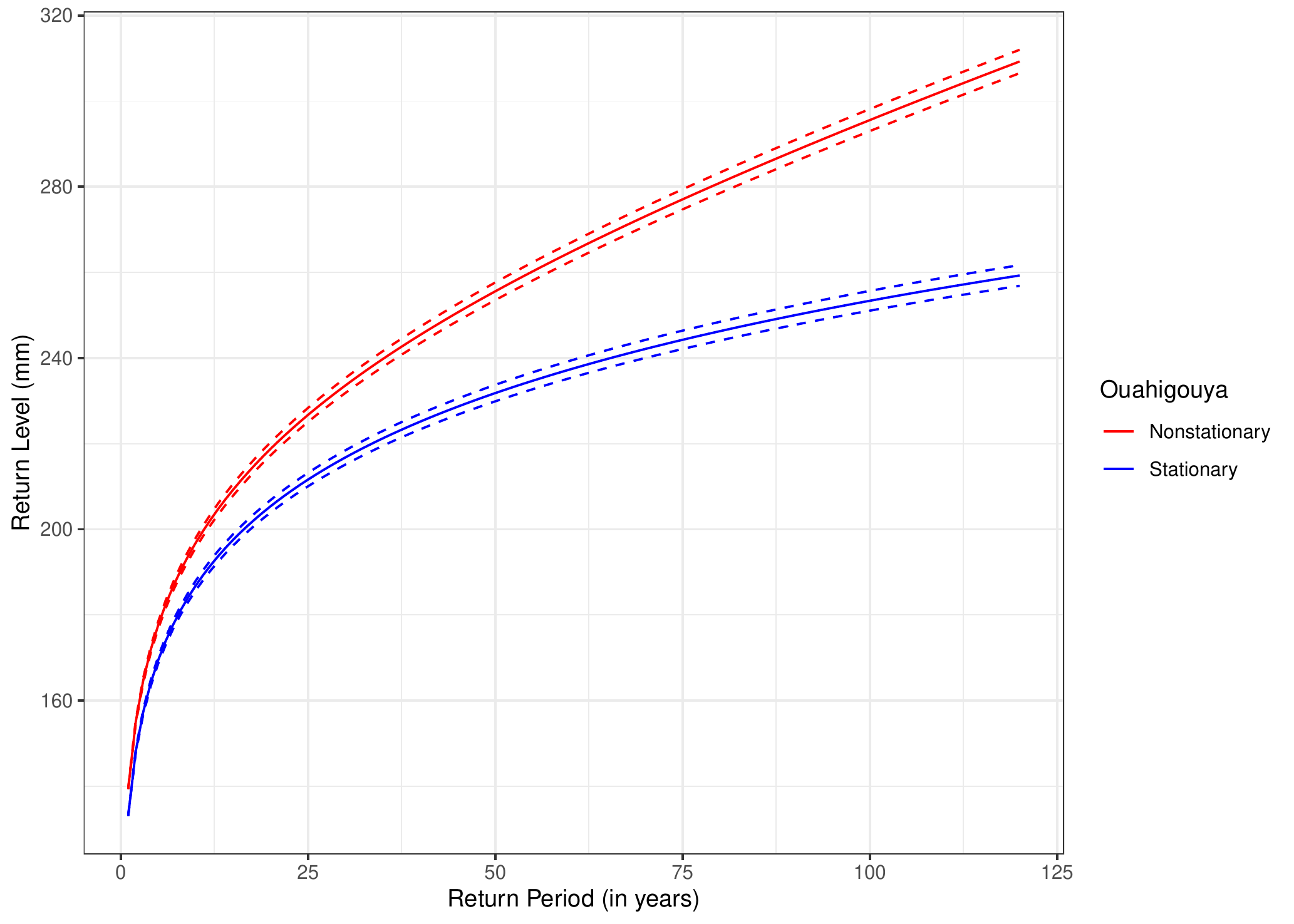}\\
   \hline
\end{tabular}
\caption{Trends of stationary and non-stationary return levels at reference stations computed with a log-linear tail trend function.}\label{figTrendsNonStatRL}
\end{figure}
For a given return period $m$, we first compute the return levels $z_{m}$ on each point of grid from the latent spatial process $Z$ using the estimated spatial model, before deriving the non-stationary return levels $x_{m}$ using the equation \ref{eqSpatialRL} of proposition  \ref{propRLNonStatRL}. We repeat this operation for different values of the return period $m$ and we deduce on each point of the grid, the associated return level (Figure \ref{figTrendsNonStatRL}). We can then produce maps of the return levels obtained from the dependence structure and the extrapolation of the trend function for a future period $m$. Thus, the extreme precipitation likely to be observed on average at least once every 50 years (resp. 100 years), will be particularly intense in the Sudanian and Sudano-Sahelian zone and less intense in the Sahelian zone, with a potentially quite strong spatial dependence within a radius of 200 km. The southwest and eastern regions of the country will be most affected by extreme precipitations. The results of the return level  for a return period $m=50$ and $100$ years are displayed in Figures (\ref{figNonStatRLsim}).

\begin{figure}[ht!!!!!!!]
\centering
\includegraphics[scale=0.26]{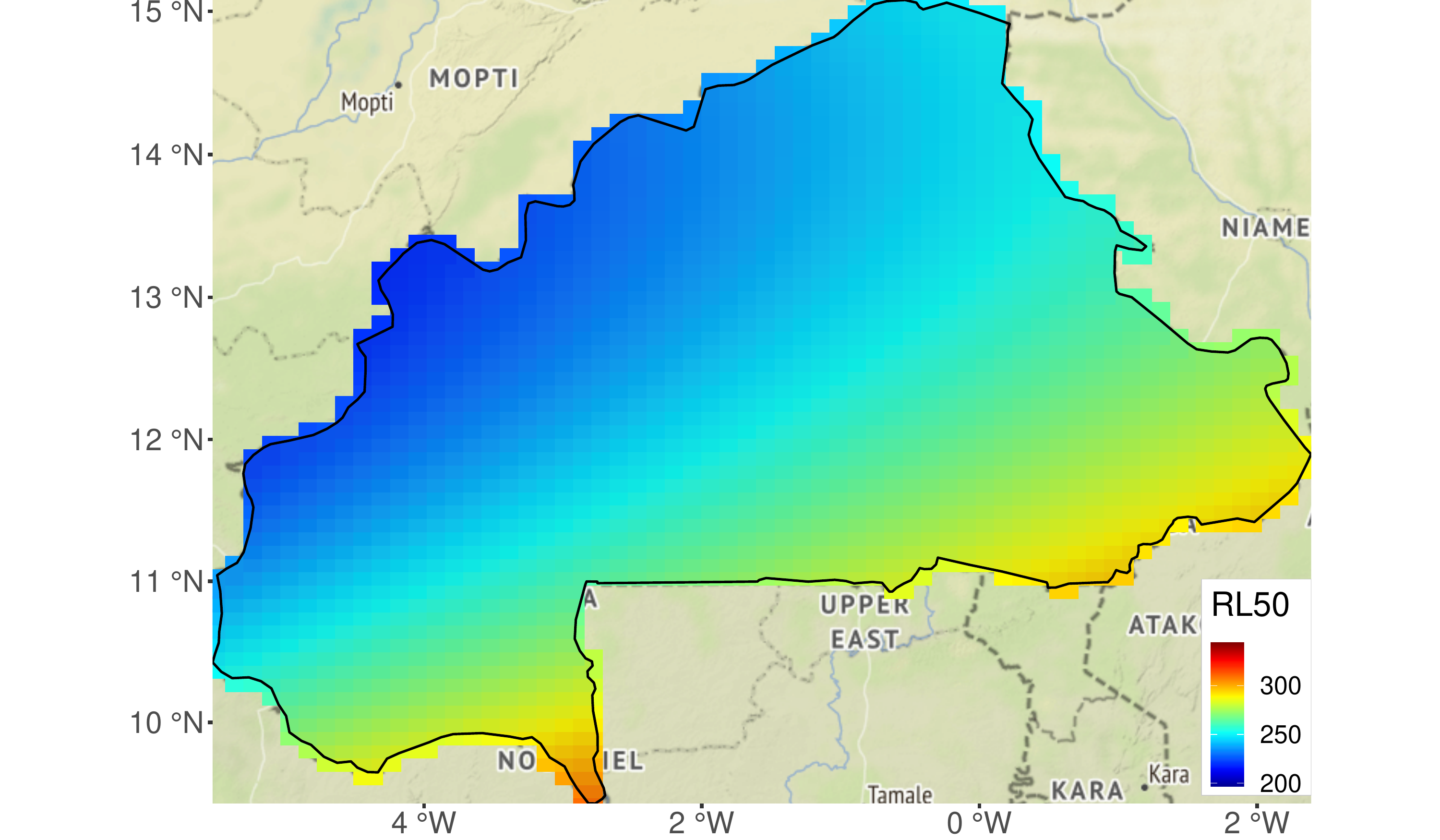}
\includegraphics[scale=0.26]{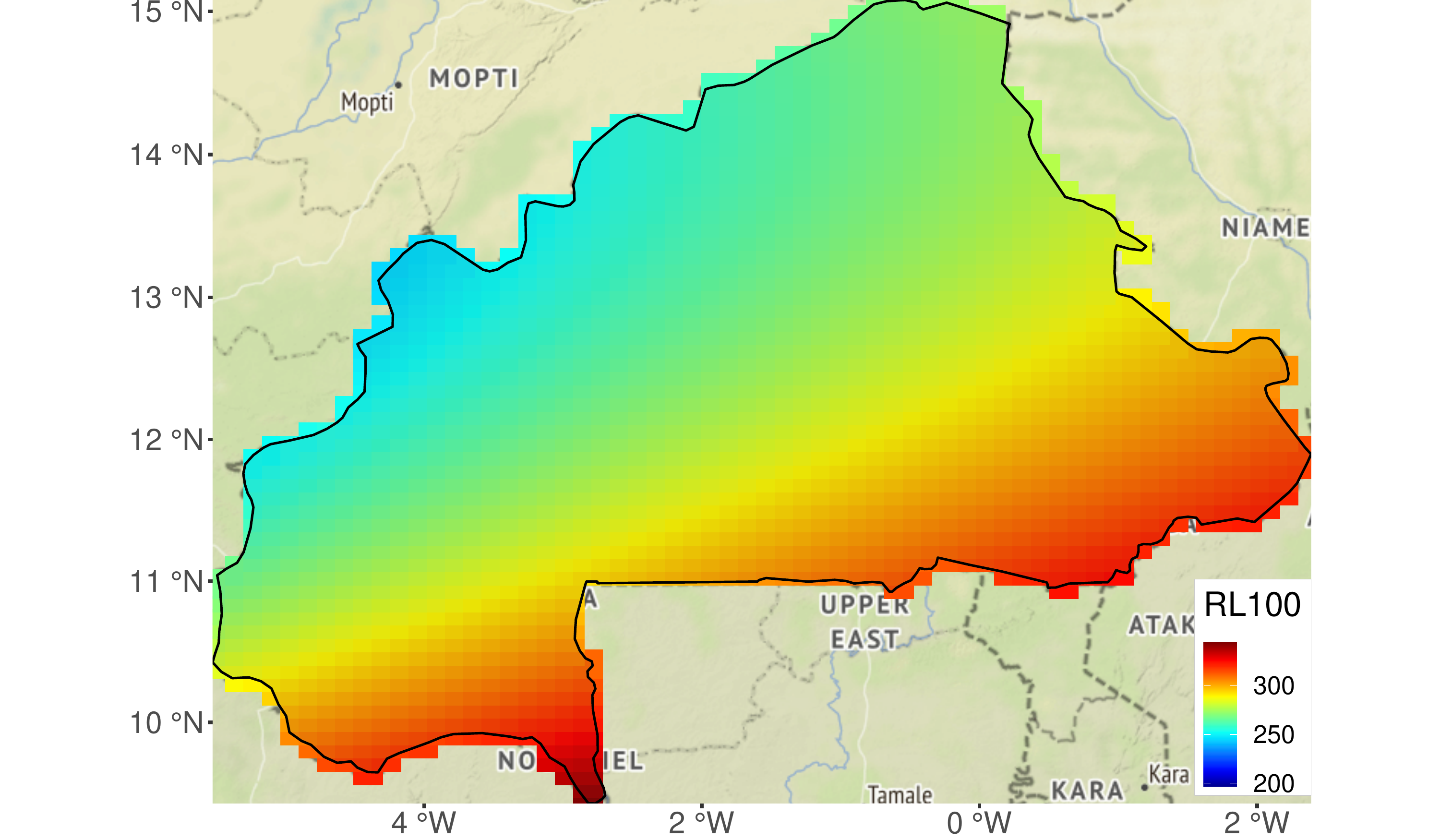}
\caption{Maps of the non-stationary 50-years return levels(left) and 100-years return level (right)  obtained by extrapolating the information from the log-linear tail trend function and the dependence structure.}\label{figNonStatRLsim}
\end{figure}

\section{Conclusion}
 In this study we proposed a new flexible methodology for trend detection in the extremes capable of capturing marginal non-stationarities and the dependence structure between margins using generalized $\ell$-Pareto processes. We computed the non-stationary return levels of rainfall in Burkina Faso at points where we have no observations and we showed that these extreme rainfall events are spatially  correlated around a radius of 200km. This spatial dependence decreases progressively as the distance increases. In sum, we set up a non-stationary stochastic generator of extreme rainfall in Burkina Faso.

%----------------------------------------------------------------------------------------------------------------
\begin{appendix}
\section{Proof of proposition \ref{NewPropReturnLevel}} 
\begin{enumerate}
\item[i)] 
\begin{proof}
Let $N_{m}$ be a  random variable describing the number of exceedances for a period of time $m$; it comes that
\begin{eqnarray} 
N_{m}= \displaystyle\sum_{t=t_{0}}^{to+n_{x}m}\mathbb{1}_{\left\lbrace X_{t}(s)> x_m(s)\right\rbrace } 
\Rightarrow E\left[ N_{m}\right] = \sum_{t=t_{0}}^{t_{0}+n_{x}m}E\left( \mathbb{1}_{\left\lbrace X_{t}(s)> x_{m}(s)) \right\rbrace }\right). \nonumber
\end{eqnarray}
And according to the equation (\ref{eqLienXandZ}), we have
\begin{eqnarray}
E\left[ N_{m}\right]  = \sum_{t=t_{0}}^{t_{0}+n_{x}m}P\left( X_{t}(s)> x_m(s)\right)= \sum_{t=t_{0}}^{t_{0}+n_{x}m}c_{\theta}\left(\frac{t}{n},s\right)P\left(Z_t(s) > x_m(s)\right). \nonumber
\end{eqnarray}
Moreover, given a sufficiently large $u(s)$ threshold with $s\in S$, 
\begin{eqnarray} 
P\left[ Z_t(s)>x_m(s)\right] &=&P\left[ Z_t(s)>u(s)\right] *P\left[ Z_t(s)-u(s)>x_m(s)\mid Z_t(s)>u(s)\right]  \nonumber\\
&\approx& \phi_{u}(s)*\bar{F}_Z\left[ x_m(s)-u(s),\sigma(s),\gamma(s)\right] , x_m(s) > u(s) \nonumber
\end{eqnarray}
Furthermore, it follows that
\begin{eqnarray}
E\left[ N_{m}\right] =1 &\Rightarrow & \displaystyle\sum_{t=t_{0}}^{t_{0}+n_{x}m}\left\lbrace c_{\theta}\left(\frac{t}{n},s\right)\phi_{u}(s)*F_Z\left(x_m(s)-u(s),\sigma(s),\gamma(s)  \right)\right\rbrace =1 \nonumber
\end{eqnarray}
Which gives 
\begin{eqnarray}
 \displaystyle\sum_{t=t_{0}}^{t_{0}+n_{x}m}\left\lbrace \phi_{u}(s) c_{\theta}\left(\frac{t}{n},s\right)\left( 1+\frac{\gamma(s)}{\sigma(s)}\{x_m(s)-u(s)\}\right)^{-1/\gamma(s)}\right\rbrace =1\nonumber
\end{eqnarray}
with $\sigma(s)=a_{n}(s)+\gamma(s)(u(s)-b_{n}(s))$
\item[ii)]  The result (\ref{eqNewEWT}) is shown in a similar way
\end{proof}
\end{enumerate}

\section{Asymptotic distribution of $\ell$-Pareto Process}
For a threshold vector $u=(u_{1},\cdots, u_{d})\in\mathbb{R}^{d}_{+}$ and for any regularly varying stochastic process $Z\in GRV\left\lbrace \gamma, a_{n}, b_{n}, \Lambda \right\rbrace $, the density function of the $\ell$-excess of a $\ell$-Pareto process is obtained by renormalizing suitably the intensity function  $\lambda $ found by taking the partial derivatives of the previously defined measure $\Lambda$  (\ref{eqhypoGRV})  by $\Lambda\left( A_{\ell}(u)\right) $:
\begin{eqnarray}\label{eqDensity}
f_{\ell,u}(z)=\dfrac{\lambda(z)}{\Lambda\left\lbrace A_{\ell}(u) \right\rbrace },  ~~ z\in A_{\ell}(u),
\end{eqnarray}
where
\begin{eqnarray}
\Lambda\left\lbrace A_{\ell}(u)\right\rbrace =\int_{A_{\ell}(u)}\lambda(z)dz,  \nonumber
\end{eqnarray} while  $\lambda$ is the intensity function and $A_{\ell}(u)=\left\lbrace z\in \mathbb{R}_{+}^{d}: \ell\left(z/u \right)\geq 1\right\rbrace $ the region of exceedances. In order to make inferences and to model the dependence structure of $\ell$-Pareto processes, we focus on the Brown-Resnick model for which the formulas of $\Lambda$ and $\lambda$ are available. The $d$-dimensional intensity function of the Brown-Resnick model is given:
\begin{eqnarray}\label{eqIntensity}
\lambda_{BR}(z)=\dfrac{\mid\Sigma\mid^{-1/2}}{z_{1}^{2}z_{2}, \cdots,z_{d}\left( 2\pi\right)^{(d-1)/2} }\exp\left(-\frac{1}{2}\bar{z}^{T}\Sigma^{-1}\bar{z} \right),~~ z\in\mathbb{R}^{d}_{+}\nonumber 
\end{eqnarray}
where $\bar{z}_{i}=\log\frac{z_{i}}{z_{1}}+\gamma(s_{i}-s_{1})$ and  $\Sigma$ the covariance matrix. 

To better capture the possible dependence structure we use anisotropic semi-variograms whose parameters change with time and other covariates. The different parameters are estimated using the gradient scoring rule method  \cite{DeFondeville2018}. For any $z\in A_{\ell}(u)$, the log-density function is given by:
\begin{eqnarray}
\ds\delta_{w}(\lambda_{\theta}, z) =\sum_{j=1}^{d}\left( 2w_{j}(z)\dfrac{\partial w_{j}(z)}{\partial z_{j}}\dfrac{\partial\log\lambda_{\theta}(z)}{\partial z_{j}}
+ w_{j}(z)^{2}\left[ \dfrac{\partial^{2}\log\lambda_{\theta}(z)}{\partial z_{j}^{2}} +\frac{1}{2}\left\lbrace  \dfrac{\partial\log \lambda_{\theta}(z)}{\partial z_{j}}\right\rbrace^{2}  \right]\right)
\end{eqnarray}
where $w$ is a differentiable weighting function that vanishing on the boundaries of $A_{\ell}(u)$ and $\theta$ is an element in the $\Theta\subset\mathbb{R}^{p}$ parameter space.
\end{appendix}

\bibliographystyle{chicago}
\bibliography{references}
\end{document}